\documentclass{article}
\usepackage{arxiv}
\usepackage[english]{babel}
\usepackage[utf8]{inputenc} 
\usepackage[T1]{fontenc}    
\usepackage{hyperref}       
\usepackage{url}            
\usepackage{booktabs}       
\usepackage{amsfonts}       
\usepackage{nicefrac}       
\usepackage{microtype}      
\usepackage{natbib}         
\usepackage{lipsum}
\usepackage{graphicx}
\graphicspath{ {./images/} }
\usepackage{xcolor}
\usepackage{dirtytalk}
\usepackage{float}
\usepackage{enumitem}
\usepackage{amsmath}
\usepackage{subcaption}
\usepackage[section]{placeins}
\usepackage[normalem]{ulem}
\usepackage{tikz}
\usetikzlibrary{positioning, shapes.geometric, calc}
\usepackage{algorithm}
\usepackage{algpseudocode}
\usepackage{appendix}
\usepackage{rotating} 

\makeatletter
\AtBeginDocument{%
  \expandafter\renewcommand\expandafter\subsection\expandafter{%
    \expandafter\@fb@secFB\subsection
  }%
}
\makeatother

\title{Macro-aware time series forecasting via hierarchical mixed-frequency attention models}

\author{
 Daniel Cunha Oliveira\\
  IME-USP \\
  \texttt{doliveira@ime.usp.br} \\
  \And
 Kieran Wood\\
  University of Oxford \\
  \texttt{kieran.wood@eng.ox.ac.uk} \\
  \And
 Stefan Zohren\\
  University of Oxford \\
  \texttt{stefan.zohren@eng.ox.ac.uk} \\
  \And
 Mihai Cucuringu\\
  UCLA \\
  University of Oxford \\
  \texttt{mihai@math.ucla.edu} \\
  \And
 André Fujita\\
  IME-USP \\
  Kyushu University \\
  \texttt{andre.fujita@ime.usp.br} \\
}

\begin{document}

\maketitle

\begin{abstract}

Deep learning models show promise in financial forecasting, yet their generalization is often undermined by small datasets, noisy signals, and non-stationarity. While meta-learning and related techniques mitigate some of these issues, they typically do not account for a core limitation in macro-financial prediction: the scarcity of distinct macroeconomic regimes that drive asset returns. We introduce HANET (Hierarchical Attention Network), a hybrid LSTM-based architecture that integrates macroeconomic domain knowledge through attention over long-run macro contexts while preserving high-frequency market dynamics. HANET organizes information in a hierarchical mixed-frequency structure, with daily asset-return signals nested within monthly macroeconomic windows, and introduces a Hierarchical Cross-Attention mechanism that reconciles low-frequency macro signals with high-frequency returns without discarding granular daily information. By framing regime selection as attention over macroeconomic contexts, the model adapts to scarce and shifting regimes. Empirically, across 55 liquid futures spanning multiple asset classes, HANET consistently outperforms neural forecasters that ignore macroeconomic information, particularly during turbulent periods, improving risk-adjusted returns and mitigating losses. Ablation studies show that these gains rely on structured macro conditioning rather than naive feature augmentation: an LSTM with the same macro representation performs poorly, and shuffling macro contexts substantially degrades performance. Finally, HANET provides interpretability through attention weights, highlighting which historical regimes are most influential for each forecast and linking macro conditions to portfolio outcomes. These results establish HANET as a systematic approach to integrating macroeconomic information into attention-based deep learning for financial forecasting.

\end{abstract}

\section{Introduction}


 




Deep learning has achieved remarkable success across many domains, yet its generalization can degrade sharply when training data is limited. Small datasets increase the risk of learning spurious patterns, a phenomenon closely tied to imbalanced and noisy samples \cite{zhang2017rethinking, he2008learning}. In financial forecasting, this challenge is amplified by non-stationarity and structural change, including time-varying predictability of stocks and bonds \cite{fama1989business, ilmanen1995time}, momentum crashes \cite{daniel2016momentum}, carry crashes \cite{brunnermeier-nagel-2008}, and correlation breaks \cite{molenaar2024stockbond}. Robust methods are therefore essential, especially in realistic multi-asset settings where models must generalize across heterogeneous instruments.

A large body of machine learning research addresses biased or scarce data via resampling (e.g., SMOTE \cite{chawla2002smote}), importance weighting \cite{kahn1953importance}, hard example mining \cite{freund1997boosting, lin2017focal}, and meta-learning \cite{ren2018learning, shu2019metaweightnet}. A complementary perspective is one-shot or few-shot learning: rather than correcting bias, the goal is to learn from a small set of relevant examples. This framing is natural for macro-finance, where long histories do not imply many independent training ``episodes'': the number of distinct macroeconomic regimes (recessions, inflation cycles, and structural shifts) is inherently scarce. Even in broad universes, such as our dataset of 55 liquid futures spanning multiple asset classes, regime diversity remains limited and non-stationarity manifests differently across sectors.

Attention over a context set, as introduced by Matching Networks \cite{vinyals-etal-2016} and further developed by Neural Processes \cite{kim-mnih-schwarz-2019}, provides a mechanism for test-time adaptation by retrieving and weighting the most relevant past contexts. Beyond performance, attention can also support interpretability: the \textit{Momentum Transformer} \citep{wood2021trading} demonstrates how attention mechanisms can be used to visualize the specific historical time steps 
associated with 
trend predictions, effectively treating attention weights as a proxy for importance.

In financial deep learning, \cite{wood2024fewshot} proposes a few-shot approach inspired by Neural Processes where attention is defined over the cross-section of assets, improving portfolio metrics in difficult post-2017 periods. While promising, this line of work leaves open a key question: how can high-dimensional \emph{external} information be integrated into attention, particularly macroeconomic variables that are central in asset pricing theory \cite{lucas1978asset, breeden-1979, epsteinzin1989} and empirics \cite{fama1989business, chen1986economic, lettau2001consumption, ang2006cross}? Moreover, macro data are typically low-frequency, creating a mixed-frequency challenge: a practical architecture must reconcile monthly macro signals with daily returns without discarding daily trading information.

To bridge this gap, we introduce HANET (Hierarchical Attention Network), a hybrid LSTM-based architecture grounded in macroeconomic domain knowledge. HANET extends few-shot attention frameworks \cite{vinyals-etal-2016, wood2024fewshot} with a \emph{Hierarchical Cross-Attention} mechanism  designed for mixed-frequency inputs. Queries and keys operate at the monthly frequency to identify relevant historical macro regimes, while values preserve granular daily market information. Monthly attention weights are then 
projected 
to the daily grid to reweight high-frequency 
representations
aligning regime information with daily returns without losing day-to-day signal structure. We evaluate HANET in a realistic multi-asset setting using 55 futures across multiple asset classes and show that macro-conditioned attention improves robustness and performance, particularly in turbulent periods where standard neural forecasters often fail. Crucially, the macro-attention mechanism also provides an interpretable lens into the model's behavior, revealing which historical regimes and macro features most strongly associate with each forecast and linking macro conditions to portfolio outcomes.

We summarize our contributions as follows:

\begin{itemize}
    \item \textbf{Macro-conditioned forecasting.}
    We propose HANET, a hybrid LSTM-based forecasting architecture that conditions predictions on high-dimensional macroeconomic contexts, enabling adaptation to scarce and shifting regimes without requiring explicit regime labels.

    \item \textbf{A Hierarchical Mixed-Frequency Attention Primitive.}
    We introduce a Hierarchical Cross-Attention mechanism that bridges disparate data frequencies by computing attention on a monthly macro grid and projecting the resulting weights onto a daily trading grid, thereby anchoring decisions in low-frequency fundamentals while preserving high-frequency signal integrity.

    \item \textbf{General-purpose systematic forecasting, including carry.}
    We provide, to our knowledge, the first deep learning formulation of the time-series carry task and demonstrate that HANET generalizes beyond momentum-style applications by capturing the relationships between macro regimes and term-structure-driven signals across assets.

    \item \textbf{Interpretability via hierarchical attention on the macro space.}
    We show how HANET's attention over macro contexts can be used to identify influential historical regimes and macroeconomic 
    conditions associated with 
    each prediction, offering an interpretable connection between macro conditions, trading signals, and portfolio performance.
\end{itemize}

\section{Related Literature}
\label{sec:related_lit}

The literature on systematic time-series strategies is extensive, with substantial evidence that momentum and carry signals capture persistent risk premia across asset classes. \citet{moskowitz-etal-2012} established time-series momentum as a pervasive phenomenon across equities, bonds, commodities, and currencies, showing that an asset's own past excess return predicts its future excess return at horizons ranging from one to twelve months. \citet{baz-etal-2015} provided a unified framework for constructing momentum and carry signals across asset classes, including the MACD-based trend filters and carry definitions we adopt as benchmarks. Carry strategies, in particular, have been studied extensively in foreign exchange by \citet{lustig-etal-2011} and generalized to a multi-asset setting by \citet{koijen-etal-2018}, who showed that carry earns a significant and persistent premium in virtually every asset class studied. Despite their strong unconditional performance, both factors exhibit episodic crash risk: \citet{daniel2016momentum} documented sharp momentum crashes following market rebounds, and \citet{brunnermeier-nagel-2008} highlighted analogous carry unwinds during periods of funding-liquidity stress. These non-stationarities motivate models that can adapt to changing market regimes rather than relying on fixed signal-to-position mappings.

Recently, due to advancements in computational power, theoretical developments in machine learning, and the availability of large financial datasets, researchers have started to apply deep learning to systematic trading. \citet{lim2019enhancing} proposed the Deep Momentum Network (DMN), which uses an LSTM to jointly learn the trend signal and the position sizing from volatility-adjusted inputs, trained end-to-end under a Sharpe-based objective. \citet{wood2021trading} extended this line of work to attention-based architectures by proposing the Momentum Transformer, which uses self-attention over price histories to capture regime shifts and simultaneously offers interpretability through attention weights. Related work has applied deep learning to limit order book modeling \citep{zhang2019deeplob}, cross-sectional return forecasting \citep{gu2020empirical}, and end-to-end portfolio construction, including graph-based approaches such as DeePM \citep{wood2026deepm}. However, a common thread in this literature is that conditioning information is drawn almost exclusively from past prices or returns. Macroeconomic context is either ignored or, when included, added as a flat feature concatenation alongside daily price-based inputs. This is a restrictive design choice given the long-standing evidence that expected returns vary with the state of the economy, from classical equilibrium models \citep{lucas1978asset, breeden-1979, epsteinzin1989} to empirical work documenting business-cycle predictability \citep{fama1989business, chen1986economic, lettau2001consumption, ang2006cross}. An important question is therefore whether structured integration of macroeconomic information can deliver gains over return-only deep learning baselines. Our ablation results, discussed in Section~\ref{sec:tsmom-task}, show that the answer depends critically on \emph{how} macro information is incorporated: naive feature augmentation of an LSTM with the same macro factors used by HANET performs substantially worse than the macro-free baseline, whereas HANET's structured cross-attention mechanism delivers 
substantial and consistent 
improvements.

A parallel development in the broader deep learning literature is the reinterpretation of attention as a retrieval mechanism rather than merely a tool for modeling temporal dependencies. Matching Networks \citep{vinyals-etal-2016} cast prediction as differentiable attention over a labeled support set, providing a natural framework for few-shot learning, and Neural Processes \citep{kim-mnih-schwarz-2019} further develop this idea with a Bayesian interpretation of test-time adaptation. In finance, \citet{wood2024fewshot} adapted this perspective by treating distinct market regimes as few-shot episodes: their Cross-Attentive Time-Series Trend Network (X-Trend) retrieves historically similar regimes at inference time, improving portfolio performance during post-2017 turbulent periods. Our work builds directly on this retrieval interpretation but differs in two important respects. First, rather than sampling discrete context episodes from past returns, we attend over the entire low-frequency macroeconomic history, exploiting the fact that macro data are naturally 
much lower frequency and lower dimensional than daily financial data
Second, and more fundamentally, we introduce a hierarchical structure in which attention is computed at the monthly frequency but applied to daily value representations, reconciling the mismatch between the horizon at which regimes are defined and the horizon at which trading decisions are taken. To our knowledge, this mixed-frequency attention primitive has not previously been developed in the financial deep learning literature.

This design choice also connects our work to the econometric literature on mixed-frequency data, which develops methods for combining variables sampled at heterogeneous frequencies. MIDAS regressions \citep{ghysels-etal-2007} and the unrestricted variants of \citet{foroni-marcellino-schumacher-2015} parsimoniously parameterize lag polynomials across frequencies, enabling tractable estimation without temporal aggregation, while dynamic factor models with mixed-frequency extensions \citep{banbura-etal-2013} provide a closely related framework for nowcasting macro aggregates from high-frequency indicators. These methods typically operate in a linear or quasi-linear setting and, crucially, focus on forecasting macro targets from financial inputs. HANET shares the goal of reconciling heterogeneous frequencies but inverts both the direction and the setting: we use low-frequency macro information to condition high-frequency trading decisions, and we do so within a nonlinear attention-based architecture that preserves the granularity of daily signals.

Finally, it is important to contextualize our work within the literature on macroeconomic regime identification. A long tradition of work uses Markov-switching models, hidden Markov models, and clustering techniques applied to 
low-dimensional macro representations 
to produce discrete regime labels, often based on the FRED-MD database of \citet{mccracken2016fredmd}, which has become a standard input for such exercises. These discrete labels are then typically fed into a separate allocation rule. Our approach differs from this two-stage paradigm: rather than committing to a single discrete regime classification, we frame regime identification as a soft attention problem over a long macro history, learned end-to-end with the trading objective. The attention weights themselves, discussed in Section~\ref{sec:interp-macro-att}, provide an interpretable view of which historical contexts the model deems most relevant for the current macro state, without requiring the researcher to pre-specify the number of regimes or their boundaries. Taken together, these contributions position HANET as a bridge between the trend-following deep learning tradition \citep{lim2019enhancing, wood2021trading}, the few-shot attention paradigm \citep{vinyals-etal-2016, wood2024fewshot}, and the macro-finance tradition that informs our choice of conditioning variables.

\section{Time Series Factor Strategies}
\label{sec:ts_factors}

To provide a theoretical foundation for our model inputs, this section introduces the construction of asset-specific factors used to drive the forecasting process. In the financial literature, factors represent systematic drivers of risk and return that capture persistent anomalies across different market regimes.

Fundamental to our implementation is the distinction between cross-sectional (CS) and time-series (TS) factor strategies, as discussed by \cite{baz-etal-2015}. Cross-sectional strategies are relative-value in nature: assets are ranked against each other based on a signal, and long--short allocations are formed to exploit dispersion in expected returns, often under a market-neutral constraint. In this framework, the investor is interested in the ordinal ranking of assets, going long those with the strongest signals and short those with the weakest to capture a cross-sectional risk premium.

In contrast, time-series strategies evaluate each asset in isolation. A TS approach determines the position in a specific instrument based solely on its own history (and, in our setting, its macroeconomic context). While CS strategies rely on cross-sectional dispersion of returns across a panel of assets, TS strategies depend on the persistence of returns over time for a given instrument. By focusing on the time-series risk premium, we allow the model to learn from historical periods where temporal persistence was the 
primary source of predictability. 
Consequently, we focus our analysis on two of the most prominent time-series factors in the literature: momentum and carry.

\paragraph{Notation.}
Let $\bigl\{(\boldsymbol{r}_t, \boldsymbol{x}_t)\bigl\}_{t=1}^{\infty}$ be a multivariate time series taking values in $\mathbb{R}^{m+d}$, where $\boldsymbol{r}_t = (r_{t,1},\dots,r_{t,m})' \in \mathbb{R}^{m}$ denotes the vector of (daily) asset returns for $m$ instruments, and $\boldsymbol{x}_t \in \mathbb{R}^{d}$ denotes a $d$-dimensional vector of observed covariates available at time $t$. For each asset $i \in \{1,\dots,m\}$, our forecaster produces a (possibly continuous) trading signal $S_{t,i} \in \mathbb{R}$ as a function of $\boldsymbol{x}_t$. Collecting signals across assets yields $\boldsymbol{S}_t = (S_{t,1},\dots,S_{t,m})' \in \mathbb{R}^{m}$. Finally, signals are mapped into positions via a sizing function $\phi:\mathbb{R}\to[-1,1]$, applied elementwise, so that the position in asset $i$ at time $t$ is $w_{t,i} = \phi\!\bigl(S_{t,i}\bigr)$ for $i=1,\dots,m$.

\paragraph{Volatility scaling and portfolio returns.}
Following the univariate time-series approach, we volatility-scale each asset's exposure so that each instrument contributes comparably to portfolio risk. Let $\sigma_{\text{tgt}}$ denote a target annualized volatility level (we take $\sigma_{\text{tgt}} = 15\%$ in our experiments), and let $\sigma_{t,i}$ be an ex-ante volatility estimate for asset $i$ formed at time $t$ using an exponentially weighted moving standard deviation over a 60-day window. We define the volatility-scaled position for asset $i$ at time $t$ as
\begin{equation}
\widetilde{w}_{t,i} \;=\; w_{t,i}\,\frac{\sigma_{\text{tgt}}}{\sigma_{t,i}}.
\end{equation}
The realized (daily) return of the strategy from $t$ to $t+1$ is then
\begin{equation}
R_{t+1}^{\text{TS}}
\;=\;
\frac{1}{m}\sum_{i=1}^{m} R_{t+1,i}^{\text{TS}},
\qquad
R_{t+1,i}^{\text{TS}}
\;=\;
\widetilde{w}_{t,i}\, r_{t+1,i}
\;=\;
w_{t,i}\,\frac{\sigma_{\text{tgt}}}{\sigma_{t,i}}\, r_{t+1,i},
\end{equation}
where $m$ is the number of assets. This standardization aligns risk contributions across heterogeneous markets, enabling the model to aggregate signals from diverse instruments into a coherent portfolio return. After volatility scaling, portfolio returns are aggregated using equal weights across assets.

\subsection{Time Series Momentum Signals}

To capture trends in asset returns, we implement two primary momentum signals that serve both as inputs to our neural network and as baseline indicators for benchmark strategies. Following the univariate time-series approach of \cite{moskowitz-etal-2012}, each asset's signal is computed independently from its own historical price dynamics.

Throughout, $i \in \{1,\dots,m\}$ indexes assets, while $w \in \mathcal{W}$ indexes the signal window (lookback / horizon) used to construct the momentum indicators. Let $P_{t,i}$ denote the price of asset $i$ at time $t$.

\paragraph{Moving Average Convergence Divergence (MACD).}
The first signal is the MACD, which identifies changes in the strength, direction, and momentum of a trend. As in \cite{baz-etal-2015}, for a window pair $k=(s,\ell)$ with $s<\ell$, define
\begin{equation}
\mathrm{MACD}_{t,i}^{(k)}
\;\equiv\;
\mathrm{MACD}_{t,i}^{(s,\ell)}
\;=\;
\mathrm{EMA}_{t,i}^{(s)}(P) \;-\; \mathrm{EMA}_{t,i}^{(\ell)}(P),
\end{equation}
where $\mathrm{EMA}_{t,i}^{(s)}(P)$ denotes the exponential moving average of the price series $\{P_{u,i}\}_{u\le t}$ with span $s$, evaluated at time $t$ (and analogously for span $\ell$).

Throughout the paper, we use MACD, and the other signals defined in this section, in two ways. 
First, as a benchmark time-series strategy, we construct a directional trading signal
\begin{equation}
S_{t,i}^{\mathrm{MACD}(k)} \;=\; \mathrm{sign}\!\bigl(\mathrm{MACD}_{t,i}^{(k)}\bigr),
\end{equation}
and map it into a position via the sizing function $\phi:\mathbb{R}\to[-1,1]$,
\begin{equation}
w_{t,i}^{\mathrm{MACD}(k)} \;=\; \phi\!\bigl(S_{t,i}^{\mathrm{MACD}(k)}\bigr),
\end{equation}
where in the benchmark case we set $\phi(s)=s$, so that $w_{t,i}^{\mathrm{MACD}(k)}\in\{-1,+1\}$.
Second, in our neural models we feed the signals directly as an input feature, and the network outputs portfolio weights $\boldsymbol{w}_t=(w_{t,1},\dots,w_{t,m})'$, yielding realized portfolio returns $\boldsymbol{w}_t'\boldsymbol{r}_{t+1}$.

\paragraph{Lagged Cumulative Returns (TSMOM).}
Following \cite{moskowitz-etal-2012}, we also utilize lagged cumulative returns (time-series momentum). For a lookback window $k \in \mathcal{K}$ (measured in trading days), define the cumulative return of asset $i$ over horizon $k$ as
\begin{equation}
R_{t,i}^{(k)} \;=\; \frac{P_{t,i}}{P_{t-k,i}} - 1.
\end{equation}
We construct a directional trading signal via
\begin{equation}
S_{t,i}^{\mathrm{Ret}(k)} \;=\; \mathrm{sign}\!\bigl(R_{t,i}^{(k)}\bigr),
\end{equation}
and map it into a position using the sizing function $\phi:\mathbb{R}\to[-1,1]$,
\begin{equation}
w_{t,i}^{\mathrm{Ret}(k)} \;=\; \phi\!\bigl(S_{t,i}^{\mathrm{Ret}(k)}\bigr).
\end{equation}
In the benchmark case we set $\phi(s)=s$, so that $w_{t,i}^{\mathrm{Ret}(k)}\in\{-1,+1\}$, yielding a long position when $R_{t,i}^{(k)}>0$ and a short position when $R_{t,i}^{(k)}<0$.

\subsection{Carry Signals}
Unlike momentum signals, which are constructed from past returns (or roll-adjusted prices), carry is derived from the slope of the term structure of forwards and futures, following \cite{baz-etal-2015}. Intuitively, carry measures the forward-looking compensation implied by the curve, computed from price differences between a near contract and a deferred contract (or, in FX, between spot and forward), scaled by the relevant time-to-maturity convention.
Our carry definitions follow the practical multi-asset conventions of \cite{baz-etal-2015}, allowing for consistent implementation across heterogeneous futures markets.

We define carry for FX forwards, equity index futures, and commodity futures. We exclude bond futures due to data availability: we observe the futures curve but not the underlying funding/financing components needed to recover total carry, which would restrict us to a rolldown-only proxy rather than the full carry concept.

Throughout, assets are indexed by $i\in\{1,\dots,m\}$. For futures-based assets, let $F_{t,i}^{(T)}$ denote the futures price at time $t$ with maturity (time-to-expiry) $T$ (measured in years), and let $T_1<T_2$ denote the maturities of the first and second listed contracts. For FX, let $S_{t,i}$ denote the spot exchange rate and let $F_{t,i}^{(3\mathrm{M})}$ denote the 3-month forward.

\paragraph{FX carry (3-month forward).}
For FX, we use the 3-month forward to imply annualized carry:
\begin{equation}
\mathrm{Carry}_{t,i}^{\mathrm{FX}}
\;=\;
4\left(\frac{S_{t,i}}{F_{t,i}^{(3\mathrm{M})}} - 1\right).
\end{equation}

\paragraph{Equity index futures carry (front vs.\ second).}
For equity index futures, we compute raw carry from the first two contracts as in \cite{baz-etal-2015}:
\begin{equation}
\mathrm{RawCarry}_{t,i}^{\mathrm{Eq}}
\;=\;
\frac{1}{T_2-T_1}\;
\frac{F_{t,i}^{(T_1)} - F_{t,i}^{(T_2)}}{F_{t,i}^{(T_2)}}.
\end{equation}
Because equity index futures exhibit seasonality (e.g., dividend and roll calendars), we remove a month-of-year component. Let $m(t)\in\{1,\dots,12\}$ denote the calendar month of date $t$, and let $\mathrm{MA}_{1\mathrm{Y},t}(\cdot)$ denote a 1-year moving average computed using information up to time $t$. Define the month-specific adjustment
\begin{equation}
\mathrm{Adj}_{t,i}^{\mathrm{Eq}}
\;=\;
\frac{1}{n_{t,m(t)}}\sum_{\tau \le t:\, m(\tau)=m(t)}
\Bigl(\mathrm{RawCarry}_{\tau,i}^{\mathrm{Eq}}
-
\mathrm{MA}_{1\mathrm{Y},\tau}\!\bigl(\mathrm{RawCarry}_{\tau,i}^{\mathrm{Eq}}\bigr)\Bigr),
\end{equation}
where $n_{t,m(t)}$ is the number of business days up to $t$ that fall in month $m(t)$. The seasonality-adjusted equity carry is then
\begin{equation}
\mathrm{Carry}_{t,i}^{\mathrm{Eq}}
\;=\;
\mathrm{RawCarry}_{t,i}^{\mathrm{Eq}} - \mathrm{Adj}_{t,i}^{\mathrm{Eq}}.
\end{equation}

\paragraph{Commodity futures carry (one-year-apart contracts).}
For commodities, to mitigate seasonal effects common in agricultural and energy markets, we use the front contract and the contract expiring one year later in the same calendar month \citep{baz-etal-2015}:
\begin{equation}
\mathrm{Carry}_{t,i}^{\mathrm{Com}}
\;=\;
\frac{F_{t,i}^{(T_1+1\mathrm{Y})} - F_{t,i}^{(T_1)}}{F_{t,i}^{(T_1+1\mathrm{Y})}}.
\end{equation}

\paragraph{Carry features and benchmark positions.}
Let $\mathrm{Carry}_{t,i}$ denote the carry measure appropriate for asset $i$ (FX, equity, or commodity as above). We use two carry-based features. First, a volatility-standardized carry measure
\begin{equation}
C_{t,i}^{\mathrm{CarryVol}}
\;=\;
\frac{\mathrm{Carry}_{t,i}}{\sigma_{t,i}},
\end{equation}
where $\sigma_{t,i}$ is the ex-ante volatility estimate defined in the volatility-scaling paragraph. Second, a directional carry signal
\begin{equation}
S_{t,i}^{\mathrm{Carry}}
\;=\;
\mathrm{sign}\!\bigl(\mathrm{Carry}_{t,i}\bigr).
\end{equation}
As with momentum, we use carry in two ways: (i) as an input feature for our neural models (using $\mathrm{Carry}_{t,i}$ and/or $C_{t,i}^{\mathrm{CarryVol}}$), and (ii) as a benchmark time-series strategy by mapping the directional signal into a position via the sizing function $\phi:\mathbb{R}\to[-1,1]$,
\begin{equation}
w_{t,i}^{\mathrm{Carry}}
\;=\;
\phi\!\bigl(S_{t,i}^{\mathrm{Carry}}\bigr),
\end{equation}
where in the benchmark case we set $\phi(s)=s$, so that $w_{t,i}^{\mathrm{Carry}}\in\{-1,+1\}$.

\section{Forecasting Framework}

\subsection{Benchmark Sequence Neural Network Model}
\label{sec:lstm_benchmark}

Our benchmark forecaster is a sequence model based on a standard Long Short-Term Memory (LSTM) cell, applied in a shared-parameter manner across assets. Recall that $\boldsymbol{r}_t=(r_{t,1},\dots,r_{t,m})'\in\mathbb{R}^m$ denotes daily returns for $m$ assets. At each time $t$, we construct an asset-level feature vector $\boldsymbol{x}_{t,i}\in\mathbb{R}^{d}$ (which may include both asset-specific signals and macro features replicated across assets), and we let $\boldsymbol{e}_i\in\mathbb{R}^{s}$ denote a learned embedding for asset $i$. We define the LSTM input for asset $i$ as the concatenated vector
\begin{equation}
\boldsymbol{z}_{t,i} \;=\; \bigl(\boldsymbol{x}_{t,i}',\,\boldsymbol{e}_i'\bigr)' \in \mathbb{R}^{d+s}.
\end{equation}

For each asset $i$, the LSTM maintains a hidden state $\boldsymbol{h}_{t,i}\in\mathbb{R}^{h}$ and a cell state $\boldsymbol{c}_{t,i}\in\mathbb{R}^{h}$, updated recursively via input, forget, and output gates:
\begin{align}
\boldsymbol{i}_{t,i} &= \sigma\!\bigl(\mathbf{U}_{i}\boldsymbol{z}_{t,i} + \mathbf{V}_{i}\boldsymbol{h}_{t-1,i} + \boldsymbol{b}_{i}\bigr),\\
\boldsymbol{f}_{t,i} &= \sigma\!\bigl(\mathbf{U}_{f}\boldsymbol{z}_{t,i} + \mathbf{V}_{f}\boldsymbol{h}_{t-1,i} + \boldsymbol{b}_{f}\bigr),\\
\boldsymbol{o}_{t,i} &= \sigma\!\bigl(\mathbf{U}_{o}\boldsymbol{z}_{t,i} + \mathbf{V}_{o}\boldsymbol{h}_{t-1,i} + \boldsymbol{b}_{o}\bigr),
\end{align}
where $\sigma(\cdot)$ denotes the sigmoid function, $\mathbf{U}_{\cdot}\in\mathbb{R}^{h\times(d+s)}$, $\mathbf{V}_{\cdot}\in\mathbb{R}^{h\times h}$, and $\boldsymbol{b}_{\cdot}\in\mathbb{R}^{h}$ are trainable parameters shared across assets. The candidate cell update is
\begin{equation}
\widetilde{\boldsymbol{c}}_{t,i}
=
\tanh\!\bigl(\mathbf{U}_{c}\boldsymbol{z}_{t,i} + \mathbf{V}_{c}\boldsymbol{h}_{t-1,i} + \boldsymbol{b}_{c}\bigr),
\end{equation}
and the cell/hidden states evolve as
\begin{align}
\boldsymbol{c}_{t,i} &= \boldsymbol{f}_{t,i}\odot \boldsymbol{c}_{t-1,i} \;+\; \boldsymbol{i}_{t,i}\odot \widetilde{\boldsymbol{c}}_{t,i},\\
\boldsymbol{h}_{t,i} &= \boldsymbol{o}_{t,i}\odot \tanh\!\bigl(\boldsymbol{c}_{t,i}\bigr),
\end{align}
where $\odot$ denotes element-wise multiplication.

To produce tradable outputs, we form an (unbounded) score for each asset via a linear head,
\begin{equation}
\widehat{S}_{t,i} \;=\; \boldsymbol{\beta}'\boldsymbol{h}_{t,i} + b,
\end{equation}
and map scores into portfolio weights using the same sizing function $\phi:\mathbb{R}\to[-1,1]$ introduced in Section~\ref{sec:ts_factors},
\begin{equation}
w_{t,i} \;=\; \phi\!\bigl(\widehat{S}_{t,i}\bigr), \qquad i=1,\dots,m,
\end{equation}
yielding $\boldsymbol{w}_t=(w_{t,1},\dots,w_{t,m})'\in\mathbb{R}^m$. The realized portfolio return is then
\begin{equation}
r^{\mathrm{port}}_{t+1} \;=\; \boldsymbol{w}_t'\boldsymbol{r}_{t+1},
\end{equation}
which serves as the economically meaningful quantity used by our Sharpe-based loss (and related objectives) during training.


\subsection{Attention via Matching Networks and Few-Shot Time-Series Retrieval}
\label{sec:attention_matching_wood}

Attention was originally introduced as a retrieval mechanism: given a query representation, the model selects and aggregates the most relevant elements from a small candidate set. This is the core idea in Matching Networks \citep{vinyals-etal-2016}, where predictions for a query example are formed by attending over a labeled support set, yielding a similarity-weighted combination of support representations. Wood et al. \cite{wood2024fewshot} adapt this principle to finance by treating changing market conditions as distinct regimes. Their Cross-Attentive Time-Series Trend Network (X-Trend) conditions forecasts and trading decisions on a context set of past regimes, transferring trend information from historically similar regimes to the current one; the context set plays the role of the support set, while the current market state acts as the query.

In this interpretation, the compared objects are embedded time-series episodes rather than i.i.d.\ samples: at each time $t$, each $c \in \mathcal{C}_t$ denotes a historical episode sampled from a window prior to $t$. A query embedding summarizes the current regime, keys encode regime descriptors for each episode, and values encode the market information to be transferred. Attention then implements a differentiable nearest-neighbor rule over $\mathcal{C}_t$, producing a similarity-weighted mixture of episodes and an interpretable mapping from predictions to the historical regimes that 
are most relevant for them.

Let $\boldsymbol{q}_t \in \mathbb{R}^{d_q}$ denote the query embedding at time $t$, and for each context episode $c\in\mathcal{C}_t$ let $\boldsymbol{k}_{t,c}\in\mathbb{R}^{d_k}$ and $\boldsymbol{v}_{t,c}\in\mathbb{R}^{d_v}$ denote its key and value representations, respectively. We project into a shared latent space of dimension $d_{\mathrm{att}}$ via trainable matrices $\mathbf{W}_q\in\mathbb{R}^{d_{\mathrm{att}}\times d_q}$, $\mathbf{W}_k\in\mathbb{R}^{d_{\mathrm{att}}\times d_k}$, and $\mathbf{W}_v\in\mathbb{R}^{d_{\mathrm{att}}\times d_v}$, and define the scaled dot-product attention weights
\begin{equation}
\alpha_{t,c}
\;=\;
\frac{\exp\!\left(\frac{1}{\sqrt{d_{\mathrm{att}}}}
\left\langle \mathbf{W}_q \boldsymbol{q}_t,\; \mathbf{W}_k \boldsymbol{k}_{t,c} \right\rangle \right)}
{\sum_{c'\in\mathcal{C}_t}\exp\!\left(\frac{1}{\sqrt{d_{\mathrm{att}}}}
\left\langle \mathbf{W}_q \boldsymbol{q}_t,\; \mathbf{W}_k \boldsymbol{k}_{t,c'} \right\rangle \right)}.
\end{equation}
The retrieved context summary is a similarity-weighted average of projected values,
\begin{equation}
\boldsymbol{s}_t
\;=\;
\mathrm{Att}\!\bigl(\boldsymbol{q}_t,\{\boldsymbol{k}_{t,c}\}_{c\in\mathcal{C}_t},\{\boldsymbol{v}_{t,c}\}_{c\in\mathcal{C}_t}\bigr)
\;=\;
\sum_{c\in\mathcal{C}_t} \alpha_{t,c}\,\mathbf{W}_v \boldsymbol{v}_{t,c}
\;\in\;
\mathbb{R}^{d_{\mathrm{att}}}.
\end{equation}

A key distinction in X-Trend is that attention operates after episode embeddings are computed: temporal structure within each episode is encoded by a recurrent model (an LSTM), and attention is used primarily to retrieve and blend information across episodes. This contrasts with Transformer-style sequence modeling, where attention is the main mechanism for capturing temporal dependencies (together with positional encodings) by attending over many time steps in a sequence. In the Matching Networks and X-Trend viewpoint, attention is therefore not primarily responsible for modeling temporal structure; rather, it is responsible for selecting which historical regimes 
are most relevant for the current regime, while retaining interpretability through the weights $\{\alpha_{t,c}\}_{c\in\mathcal{C}_t}$.

\begin{figure}[h]
    \centering
    \includegraphics[width=0.75\textwidth]{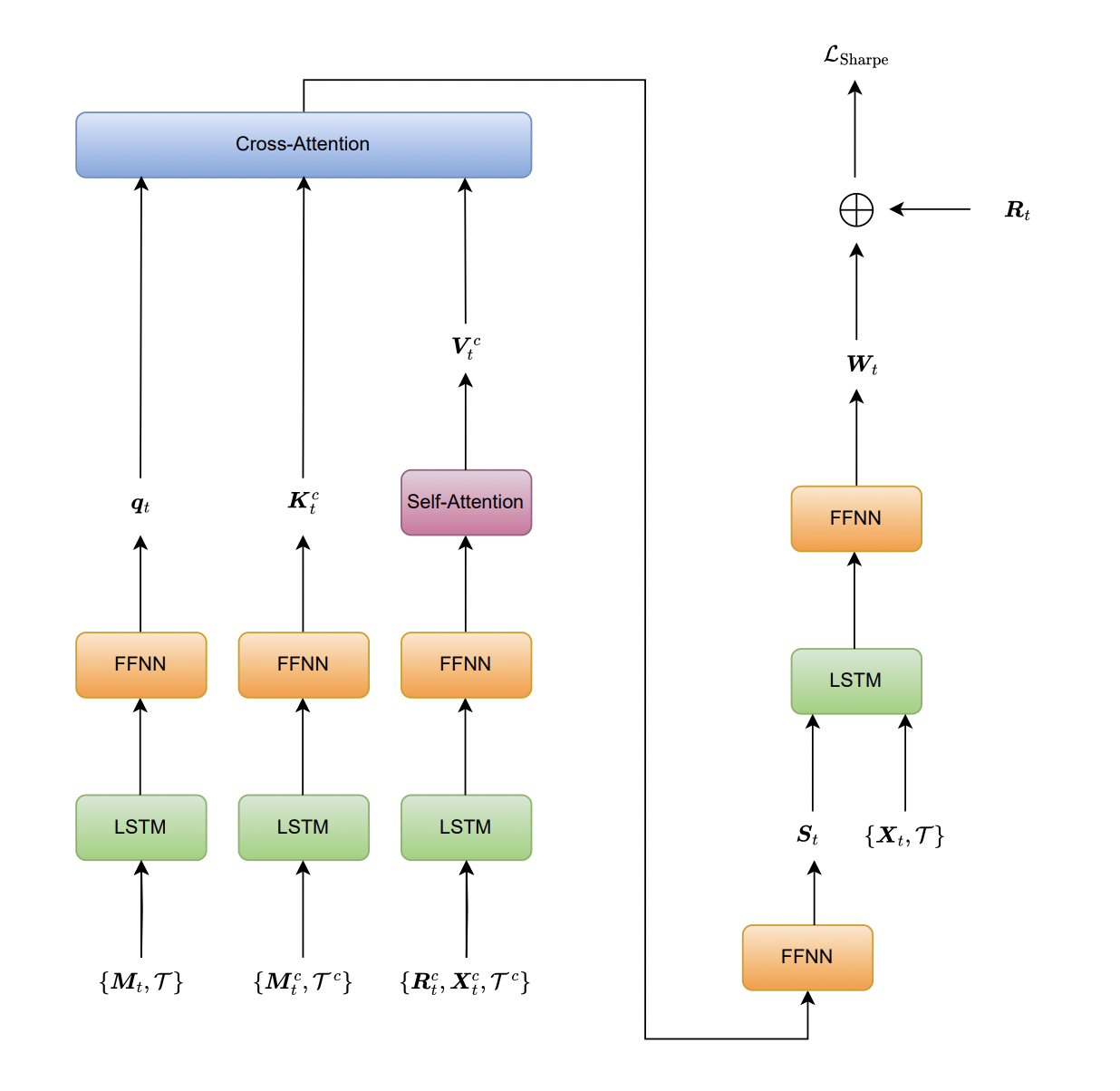}
    \caption{Diagram of the adapted Cross Attentive Time-Series Trend Network.}
    \label{fig:diag-adapted-x-trend}
\end{figure}

\subsection{Hierarchical Attention Network}
\label{sec:hanet}

Our proposed Hierarchical Attention Network (HANET) extends the retrieval-based attention paradigm of \cite{vinyals-etal-2016, wood2024fewshot} to a mixed-frequency setting. The goal is to condition high-frequency trading decisions on low-frequency macroeconomic structure in a way that is computationally efficient and economically interpretable. Rather than sampling a small context set of past episodes as in Matching Networks and X-Trend, HANET attends over the full history of a low-frequency macro sequence (monthly in our experiments, but any frequency lower than the decision horizon is admissible). Since attention cost scales with the number of keys, performing attention over a compressed low-frequency history enables conditioning on long macro histories without the prohibitive cost that would arise from attending over the full high-frequency record.

\begin{figure}[h]
  \centering
  \includegraphics[width=0.65\textwidth]{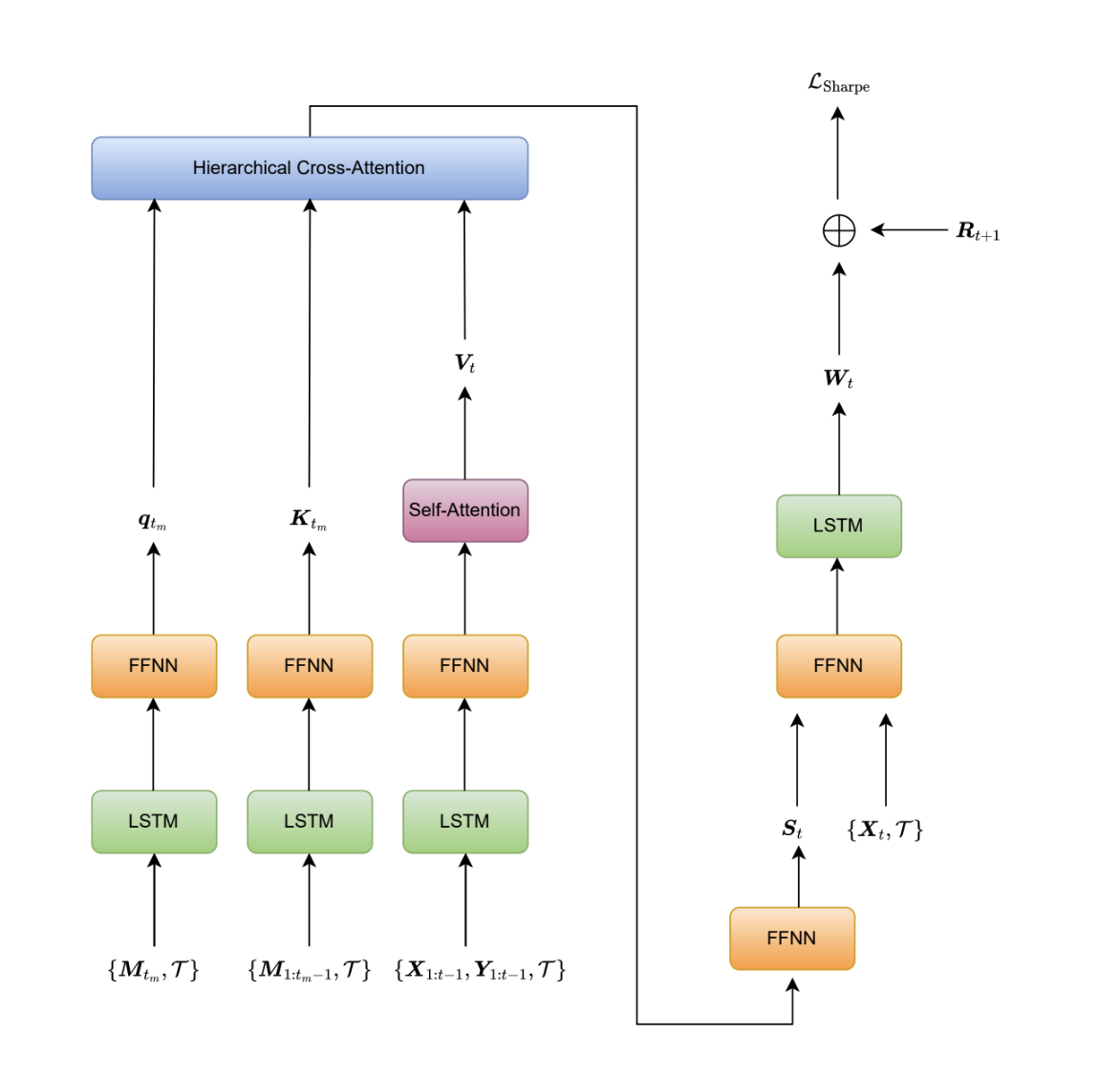}
  \caption{Diagram of the Hierarchical Attention Network (HANET) with an encoder--decoder mechanism.}
  \label{fig:diag-hanet}
\end{figure}

Figure~\ref{fig:diag-hanet} summarizes the architecture. We denote daily decision times by $t\in\{1,\dots,T\}$ and monthly macro times by $k\in\{1,\dots,K\}$. Let
\[
\mu:\{1,\dots,T\}\to\{1,\dots,K\},
\qquad
t \mapsto \mu(t),
\]
be the deterministic mapping that assigns each day $t$ to its corresponding month index.\footnote{For example, $\mu(t)$ can be defined by the calendar month label induced by $t$; all days within the same calendar month share the same $\mu(t)$.}
Let $\boldsymbol{\kappa}_k\in\mathbb{R}^{d_\kappa}$ denote the low-frequency macro vector at month $k$, and let $\boldsymbol{x}_{t,i}\in\mathbb{R}^{d}$ denote the daily covariates for asset $i$ at time $t$ (as in the notation preamble). We also allow auxiliary inputs $\mathcal{T}$ shared across modules (e.g., calendar features and/or asset identifiers). In our implementation, $\boldsymbol{\kappa}_k$ is constructed via PCA fitted on the training set: loadings are estimated using only training data and then held fixed, so that validation and test macro vectors are obtained by projection onto the training-fitted eigenvectors.

The encoder consists of three streams that mirror the query--key--value decomposition of attention but are adapted to the mixed-frequency setting.

First, the current-month macro state is encoded into a query representation. Let $\boldsymbol{\kappa}_{\mu(t)}$ denote the macro vector for the month containing day $t$. We form a query embedding
\begin{equation}
\boldsymbol{q}_{\mu(t)}
=
f_q\!\left(\mathrm{LSTM}_q\bigl(\{\boldsymbol{\kappa}_{\mu(t)},\mathcal{T}\}\bigr)\right)
\in \mathbb{R}^{d_{\mathrm{att}}}.
\label{eq:hanet_query}
\end{equation}

Second, the full macro history is encoded into a sequence of key embeddings. Let $\{\boldsymbol{\kappa}_{1:\mu(t)-1}\}$ denote the historical macro sequence up to month $\mu(t)-1$. Passing this sequence through a second LSTM and FFNN yields
\begin{equation}
\boldsymbol{k}_{1:\mu(t)-1}
=
f_k\!\left(\mathrm{LSTM}_k\bigl(\{\boldsymbol{\kappa}_{1:\mu(t)-1},\mathcal{T}\}\bigr)\right)
\in \mathbb{R}^{(\mu(t)-1)\times d_{\mathrm{att}}},
\label{eq:hanet_keys}
\end{equation}
where $\boldsymbol{k}_{\tau}\in\mathbb{R}^{d_{\mathrm{att}}}$ denotes the key embedding associated with month $\tau\le \mu(t)-1$.

Third, value representations are constructed from high-frequency market information and refined by self-attention before being combined by cross-attention. Let $L$ denote a daily lookback length. For each day $u\in\{t-L,\dots,t-1\}$, we form an asset-level input $\boldsymbol{z}_{u,i}$ (e.g., by concatenating $\boldsymbol{x}_{u,i}$ with the asset embedding and calendar controls, as in Section~\ref{sec:lstm_benchmark}). Encoding the window $\{\boldsymbol{z}_{t-L:t-1,i}\}$ through an LSTM and FFNN yields a latent sequence which is then processed by a self-attention block:
\begin{equation}
\widetilde{\mathbf{V}}_{t-L:t-1,i}
=
f_{v,1}\!\left(\mathrm{LSTM}_v\bigl(\{\boldsymbol{z}_{t-L:t-1,i}\}\bigr)\right),
\qquad
\mathbf{V}_{t-L:t-1,i}
=
f_{v,2}\!\left(\mathrm{SelfAtt}\bigl(\widetilde{\mathbf{V}}_{t-L:t-1,i}\bigr)\right),
\label{eq:hanet_values}
\end{equation}
where $\mathbf{V}_{t-L:t-1,i}$ denotes the high-frequency value sequence made available to the cross-attention module for asset $i$.

HANET then performs hierarchical cross-attention that aligns the low-frequency query--key space with the high-frequency value space. To control the sharpness of the monthly attention distribution, we introduce a learnable positive temperature parameter
\begin{equation}
\tau_{\mathrm{att}}
=
\exp(\theta_{\mathrm{att}})
>0,
\label{eq:hanet_temperature}
\end{equation}
where $\theta_{\mathrm{att}}\in\mathbb{R}$ is an unconstrained trainable scalar. Smaller values of $\tau_{\mathrm{att}}$ produce sharper attention over historical macro months, whereas larger values generate a more diffuse weighting scheme.

Attention weights are computed at the monthly level and then expanded to the daily grid. Specifically, for day $t$ we compute
\begin{equation}
\alpha_{\mu(t),\tau}
=
\frac{
\exp\!\left(
\frac{1}{\tau_{\mathrm{att}}\sqrt{d_{\mathrm{att}}}}
\left\langle
\mathbf{W}_q \boldsymbol{q}_{\mu(t)},\,
\mathbf{W}_k \boldsymbol{k}_{\tau}
\right\rangle
\right)
}{
\sum_{\ell=1}^{\mu(t)-1}
\exp\!\left(
\frac{1}{\tau_{\mathrm{att}}\sqrt{d_{\mathrm{att}}}}
\left\langle
\mathbf{W}_q \boldsymbol{q}_{\mu(t)},\,
\mathbf{W}_k \boldsymbol{k}_{\ell}
\right\rangle
\right)
},
\qquad \tau=1,\dots,\mu(t)-1,
\label{eq:hanet_att_weights}
\end{equation}
and assign each day $u$ in the high-frequency window a weight according to its month,
\[
\widetilde{\alpha}_{t,u} \;=\; \alpha_{\mu(t),\,\mu(u)}.
\]
The macro-attended summary for asset $i$ at day $t$ is then obtained by reweighting the daily value sequence:
\begin{equation}
\boldsymbol{s}_{t,i}
=
f_{\mathrm{att}}\!\left(
\sum_{u=t-L}^{t-1} \widetilde{\alpha}_{t,u}\,\mathbf{W}_v \boldsymbol{v}_{u,i}
\right)
\in \mathbb{R}^{d_{\mathrm{att}}},
\label{eq:hanet_summary}
\end{equation}
where $\boldsymbol{v}_{u,i}$ denotes the $u$-th row of $\mathbf{V}_{t-L:t-1,i}$. This construction preserves high-frequency information in the value stream while using low-frequency macro similarity weights as a gating signal. The learnable temperature in \eqref{eq:hanet_temperature} allows the model to adapt endogenously how selective this macro retrieval step should be across historical months.

The decoder integrates the macro-attended summary with contemporaneous daily predictors to produce portfolio weights. For each asset $i$, we form a fused input
\begin{equation}
\boldsymbol{u}_{t,i}
=
f_{\mathrm{fus}}\!\bigl([\boldsymbol{s}_{t,i},\,\boldsymbol{x}_{t,i},\,\mathcal{T}]\bigr),
\end{equation}
pass it through a daily LSTM decoder, and map the resulting hidden state to a tradable weight:
\begin{equation}
\boldsymbol{h}_{t,i} = \mathrm{LSTM}_{\mathrm{dec}}(\boldsymbol{u}_{t,i}),
\qquad
\widehat{S}_{t,i} = \boldsymbol{\beta}'\boldsymbol{h}_{t,i} + b,
\qquad
w_{t,i} = \phi\!\bigl(\widehat{S}_{t,i}\bigr) = \tanh\!\bigl(\widehat{S}_{t,i}\bigr),
\label{eq:hanet_weights}
\end{equation}
where in our neural models we set the sizing function to $\phi(s)=\tanh(s)$ so that $w_{t,i}\in(-1,1)$. Collecting weights yields $\boldsymbol{w}_t=(w_{t,1},\dots,w_{t,m})'$, and the realized portfolio return is $\boldsymbol{w}_t'\boldsymbol{r}_{t+1}$. By construction, the attention weights $\{\alpha_{\mu(t),\tau}\}_{\tau\le \mu(t)-1}$ provide a low-frequency interpretability layer: they quantify which historical macro months the model deems most relevant when producing daily portfolio decisions.

\section{Dataset}

In this section, we describe the dataset used in our experiments. Section \ref{sec:financial-data} begins with the financial data, which consists of the information required to construct returns and signals for both the time-series momentum and carry tasks. We also detail the transformations applied to prices and returns in order to convert them into tradable signals for each task. Section \ref{sec:economic-data} then introduces the macroeconomic database we employ, along with the regime detection model whose outputs serve as inputs to our forecasting framework.

\subsection{Financial Data}\label{sec:financial-data}
Our financial dataset is constructed from futures and related market series gathered from the Pinnacle database. We select 55 contracts spanning commodities, bonds, currencies, and equity indices from the larger universe covered by Pinnacle; the complete list of instruments used in our experiments (by Pinnacle ticker, contract name, and exchange) is reported in Appendix~\ref{app:dataset_details}.

For all futures-based assets, we work with daily data fields (open, high, low, close/settle, volume, and open interest). In particular, Pinnacle defines the ``close'' as the exchange settle price, which we treat as the end-of-day reference for return construction. This choice ensures that all features used at time $t$ can be formed from information available up to the end of day $t$.

For both the trend (momentum) and carry tasks, we use continuously linked futures time series. Pinnacle stores the raw master commodity history in internal master files that are not directly readable, and provides the utility \texttt{DMAKERW} to extract and build continuously linked series in standard formats. We use the ratio-adjusted linked series (``.RAD'') as our baseline continuous price input for both the momentum and carry tasks, which splices prior contracts by multiplicatively scaling them to remove inter-contract gaps while preserving percentage returns. Ratio adjustment is particularly well-suited to our setting because it keeps return magnitudes consistent across roll dates, which is essential both for trend signal construction and for the carry computations that rely on clean relative price differences along the term structure.

For carry proxies, we additionally rely on term-structure information beyond the single continuous series, since carry is defined from the slope of the term structure. Accordingly, for commodity and equity index futures we use the futures curve (near and deferred maturities) to compute the carry measures described in Section~\ref{sec:financial-data} (and formalized in Section~3.1). Concretely, equity carry uses the front and second contracts, while commodity carry uses the front contract and the contract expiring one year later (same calendar month) to mitigate seasonality effects.

For FX, we combine Pinnacle's spot FX series with short-horizon forward-looking information. Pinnacle distributes FX prices and corresponding FX interest-rate series in dedicated directories, which we use to construct (or proxy) the 3-month forward needed for FX carry. In practice, we treat the resulting 3-month forward proxy as the $F_{t,3\text{M}}$ input to the carry definition.

Finally, we note that while our asset universe includes bond futures for the momentum task (via continuous series; Appendix~\ref{app:dataset_details}), we exclude bond futures from the carry experiments because we only observe the futures curve in our data environment, which would restrict us to a rolldown-only proxy rather than the full carry concept.

\subsection{Economic Data}\label{sec:economic-data}
We employ two complementary representations of the macroeconomic dataset: (i) the raw version of FRED-MD, and (ii) a latent, dimension-reduced representation constructed via principal component analysis (PCA). This dual representation allows us to (a) feed the model with direct macroeconomic inputs in an end-to-end manner and (b) work with a compact set of latent macro factors that summarizes the common variation in the macro panel and is especially convenient for attention over long histories.

The raw dataset is constructed from FRED-MD, the monthly U.S.\ macroeconomic database introduced by \cite{mccracken2016fredmd} and curated by the Federal Reserve Bank of St.\ Louis. FRED-MD is publicly available, widely used in empirical economics, and has become a standard benchmark in forecasting applications.

Importantly, 
our implementation uses point-in-time macroeconomic vintages aligned with information available at each forecast date, accounting for publication lags and data revisions, making it particularly suitable  
for backtesting and pseudo-out-of-sample forecasting experiments. 
Each monthly vintage reflects the macroeconomic information that was actually available to market participants at the time of release, accounting for publication lags and subsequent data revisions. This property is essential for our empirical setting: it ensures that the macro inputs fed to HANET at any query date $t_m$ correspond to information that was genuinely observable at $t_m$, thereby eliminating look-ahead bias from revised or retroactively updated macro series and making the reported out-of-sample performance a credible proxy for what the model could have achieved in live deployment.

Our version of the dataset contains 127 variables covering the following categories:
\begin{enumerate}
    \item Output and Income,
    \item Consumption, Orders, and Inventories,
    \item Labor Market,
    \item Housing,
    \item Money and Credit,
    \item Prices.
\end{enumerate}
To maintain a focus on U.S.-specific macroeconomic drivers and avoid redundancy with financial covariates already included in the forecasting tasks, we exclude interest rate and exchange rate variables. The sample spans December 1959 to January 2023 at a monthly frequency, with each observation representing the macroeconomic state of the economy for that month.

In addition to the raw macro panel, we construct a latent FRED-MD representation via PCA. Let 

$\boldsymbol{z}_{t_m}\in\mathbb{R}^{p}$ denote the standardized macro vector at month $t_m$, where $p$ is the number of retained indicators after the exclusions above. PCA extracts orthonormal loading vectors $\{\boldsymbol{u}_k\}_{k=1}^{K}$ that explain the dominant directions of variation in $\{\boldsymbol{z}_{t_m}\}$, and defines latent factors

\begin{equation}
\boldsymbol{M}_{t_m} = \boldsymbol{U}_K^{\top}\boldsymbol{z}_{t_m}\in\mathbb{R}^{K},
\qquad
\boldsymbol{U}_K = [\boldsymbol{u}_1,\dots,\boldsymbol{u}_K]\in\mathbb{R}^{p\times K}.
\end{equation}

Crucially, to avoid leakage, the PCA loadings $\boldsymbol{U}_K$ are fit using training data only and then held fixed: for validation and test months, $\boldsymbol{M}_{t_m}$ is computed by projecting the corresponding macro observations onto the training-fitted eigenvectors. This produces a stable low-dimensional macro representation that can be used as the monthly sequence $\{\boldsymbol{M}_{t_m}\}$ in HANET, enabling attention over the full macro history with substantially reduced computational cost relative to attending over the full raw macro panel.

\section{Performance Evaluation}

Portfolio construction is performed as follows. At each time $t$, the network outputs a matrix of raw positions 
\(\boldsymbol{W}_t \in \mathbb{R}^{n \times m}\), where each entry \(w_{t,j,i}\) corresponds to the position for asset \(i\) in sample \(j\). These raw positions reflect the model’s directional views and relative magnitudes across assets. To ensure comparability across assets and control portfolio risk, we apply volatility targeting. Specifically, the volatility-scaled portfolio weights are defined as
\begin{equation}
\tilde{w}_{t,j,i} = w_{t,j,i}  \frac{\sigma_{\mathrm{tgt}}}{\hat{\sigma}_{t,i}},
\end{equation}
where \(\hat{\sigma}_{t,i}\) denotes the ex-ante volatility of asset \(i\), estimated from an exponentially weighted moving standard deviation of past returns, and \(\sigma_{\mathrm{tgt}}\) is a fixed annualized volatility target. In this way, the network determines the raw signals \(\boldsymbol{W}_t\), while volatility scaling helps ensure that each asset contributes approximately equal ex-ante risk to the portfolio. Transaction costs can be included as a penalty term, though in the baseline analysis we set them to zero.

The one-period portfolio return for sample \(j\) is then given by
\begin{equation}
R^{\mathrm{port}}_{t+1,j} = \tilde{\boldsymbol{w}}_{t,j}^{\prime} \boldsymbol{r}_{t+1,j},
\end{equation}
where \(\tilde{\boldsymbol{w}}_{t,j} = (\tilde{w}_{t,j,1}, \dots, \tilde{w}_{t,j,m})^{\prime}\) are the volatility-scaled weights. Cumulative returns are obtained by compounding \(R^{\mathrm{port}}_{t+1,j}\) over the evaluation horizon.

\section{Empirical Results}

\subsection{Time Series Momentum Task}
\label{sec:tsmom-task}

\begin{figure}[H]
  \centering
  \includegraphics[width=0.9\textwidth]{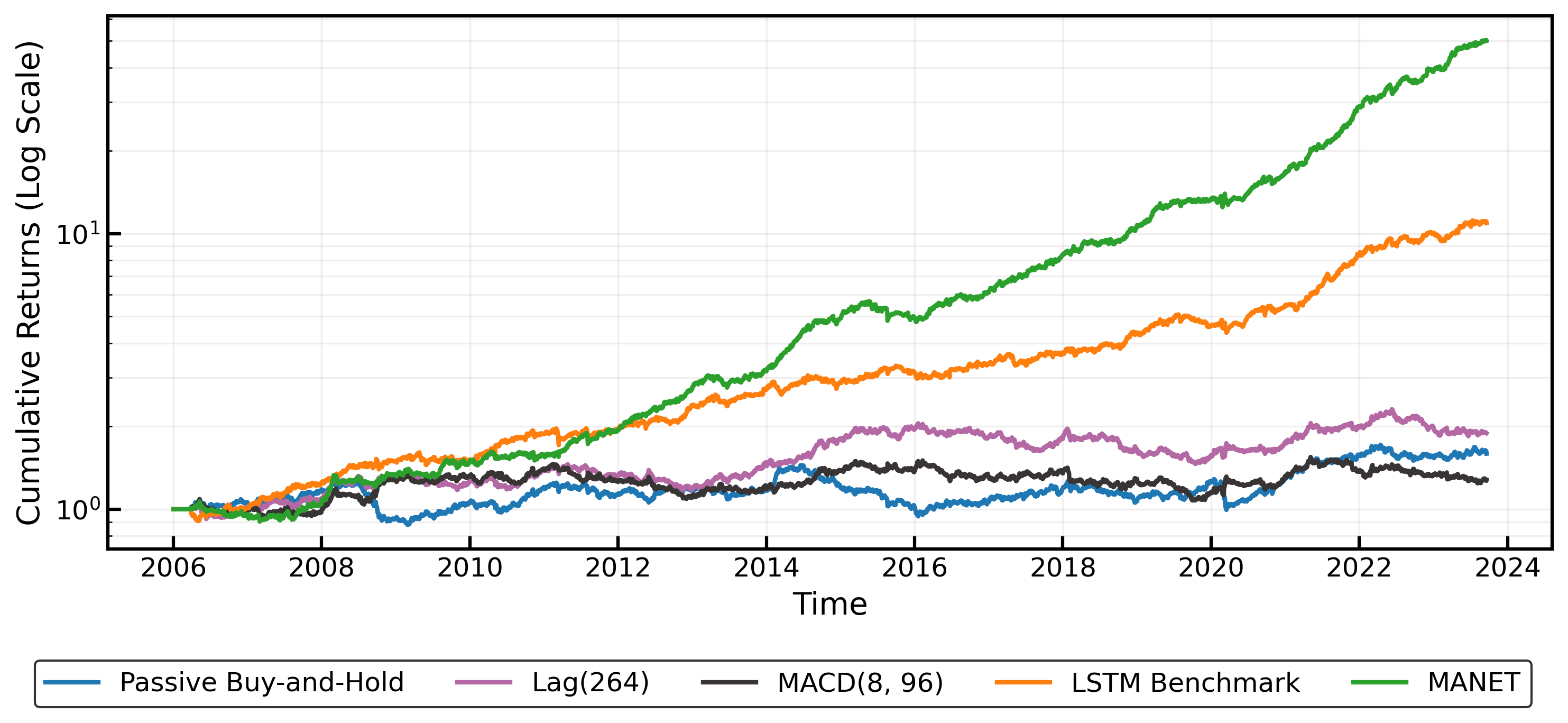}
 \caption{Cumulative log-scale returns (top panel) and drawdowns (bottom panel) for volatility-targeted (10\% annualized) time-series momentum portfolios over the 2005--2024 out-of-sample period. Strategies shown are HANET (green), the LSTM benchmark (orange), Lag(264), MACD(8,\,96), and the passive buy-and-hold benchmark. A steeper upward slope in the top panel indicates faster compounding; shallower troughs in the bottom panel indicate milder and shorter-lived drawdown episodes.}
  \label{fig:tsmom-all-cum-ret}
\end{figure}

Table~\ref{tab:tsmom-all-portfolio-metrics} and Figure~\ref{fig:tsmom-all-cum-ret} show that HANET improves the performance of standard time-series momentum portfolios in two complementary ways. First, HANET 
substantially improves a large increase in risk-adjusted returns relative to classical rules and neural baselines: at the same 10\% volatility target (see Section~\ref{sec:ts_factors}), HANET achieves an annualized mean return of 22.45\% with a Sharpe ratio of 1.99 and a Sortino ratio of 2.50, substantially exceeding both the LSTM benchmark (13.73\% mean, 1.19 Sharpe, 1.61 Sortino) and the rule-based signals: Lag(264) achieves only a 0.34 Sharpe while MACD(8, 96) delivers 0.21. Importantly, the annualized volatilities are comparable across all strategies (ranging from 10.47\% to 11.52\%), confirming that the gains are driven by improved timing and cross-asset allocation rather than by higher ex-ante 
risk exposure. 
The statistical significance of HANET's outperformance is also notable: its mean-return $t$-statistic of 8.78 is the highest in the panel, 
well above conventional significance thresholds.

Second, HANET materially improves downside outcomes compared to classical momentum rules, which exhibit severe crash risk over the sample with maximum drawdowns exceeding 28\% (and $-34.73\%$ for the passive buy-and-hold benchmark). In contrast, HANET limits its worst peak-to-trough loss to $-15.87\%$, comparable to the LSTM benchmark's best-in-class $-14.16\%$. The trade-off between HANET and the LSTM on this dimension is modest: HANET sacrifices roughly 1.7 percentage points of drawdown protection in exchange for a near-doubling of the Sharpe ratio.

Relative to the passive benchmark, HANET exhibits the highest information ratio (1.72) and the strongest active-return $t$-statistic (7.04, HAC-adjusted), confirming that the excess returns are both economically large and statistically robust. The correlation of HANET's returns with the benchmark ($\rho = 0.35$) is moderate, indicating that a meaningful share of the strategy's return comes from timing decisions that 
differ 
from a simple long-only exposure. By comparison, the LSTM benchmark has a lower correlation with the benchmark ($\rho = 0.22$) but a considerably lower information ratio (0.83), suggesting it makes more independent but less profitable bets. The classical rules offer negligible active returns: Lag(264) has an IR of just 0.05 ($t$-stat of 0.19), and MACD(8, 96) produces a negative IR.

Figure~\ref{fig:tsmom-all-cum-ret} reinforces these findings visually. On a log scale, HANET (green) compounds returns at a markedly steeper rate than all alternatives throughout the 2005--2024 sample, pulling away from the LSTM benchmark (orange) especially after 2014. The classical strategies and the passive benchmark cluster together near the bottom of the chart, with cumulative returns barely exceeding their starting value by the end of the sample. The drawdown panel further highlights that HANET's losses, though not the smallest in every episode, 
tend to recover more quickly,  
in contrast to the prolonged drawdown episodes suffered by the rule-based approaches during the 2008 financial crisis and subsequent turbulent periods. HANET's average holding period of 3 days (versus 5 for the LSTM and 34--217 for the classical rules) reflects a more active trading style that rapidly adjusts positions in response to changing conditions.

\begin{sidewaystable}
\centering
\scriptsize
\setlength{\tabcolsep}{8.0pt}
\renewcommand{\arraystretch}{1.05}
\begin{tabular}{l rrrrrrr rrr}
\toprule
 & \multicolumn{7}{c}{Gross} & \multicolumn{3}{c}{vs Bench.\ (Gross)} \\
\cmidrule(lr){2-8}\cmidrule(lr){9-11}
Strategy & Mean & Std & SR & Sortino & MDD & Hold (Days) & $t$-stat$_1$ & $\rho$ & IR & $t$-stat$_2$ \\
\midrule
Passive Buy-and-Hold (Bench.) & 2.88 & 10.47 & 0.28 & 0.37 & -34.73 & 217 & 1.28 & 1.00 & -- & -- \\
\addlinespace[1pt]
Lag(264) & 3.52 & 10.47 & 0.34 & 0.48 & -28.85 & 34 & 1.44 & 0.12 & 0.05 & 0.19 \\
MACD(8, 96) & 2.26 & 10.77 & 0.21 & 0.30 & -28.39 & 36 & 0.75 & -0.09 & -0.04 & -0.16 \\
LSTM Benchmark & 13.73 & 11.52 & 1.19 & 1.61 & \textbf{-14.16} & 5 & 5.37 & 0.22 & 0.83 & 3.44 \\
HANET & \textbf{22.45} & 11.31 & \textbf{1.99} & \textbf{2.50} & -15.87 & 3 & \textbf{8.78} & 0.35 & \textbf{1.72} & \textbf{7.04} \\
\bottomrule
\end{tabular}
\caption{Gross performance summary for volatility-targeted (10\% annualized) time-series momentum portfolios over the 2005--2024 out-of-sample period. Mean and Std denote annualized mean return and volatility (\%), SR is the annualized Sharpe ratio, Sortino is the annualized Sortino ratio, and MDD is the maximum peak-to-trough drawdown (\%). Hold (Days) reports the average number of trading days a position is maintained before reversal. The right panel reports statistics relative to the passive buy-and-hold benchmark: $\rho$ is the Pearson correlation of daily strategy returns with benchmark returns, IR is the annualized information ratio (mean active return divided by tracking error), and $t$-stat$_2$ is the Newey--West HAC $t$-statistic of active returns. Higher SR, Sortino, and IR indicate superior risk-adjusted performance; a less negative MDD indicates better drawdown control. Best values across strategies are highlighted in bold.}
\label{tab:tsmom-all-portfolio-metrics}
\end{sidewaystable}

\begin{figure}[H]
  \centering
  \includegraphics[width=0.75\textwidth]{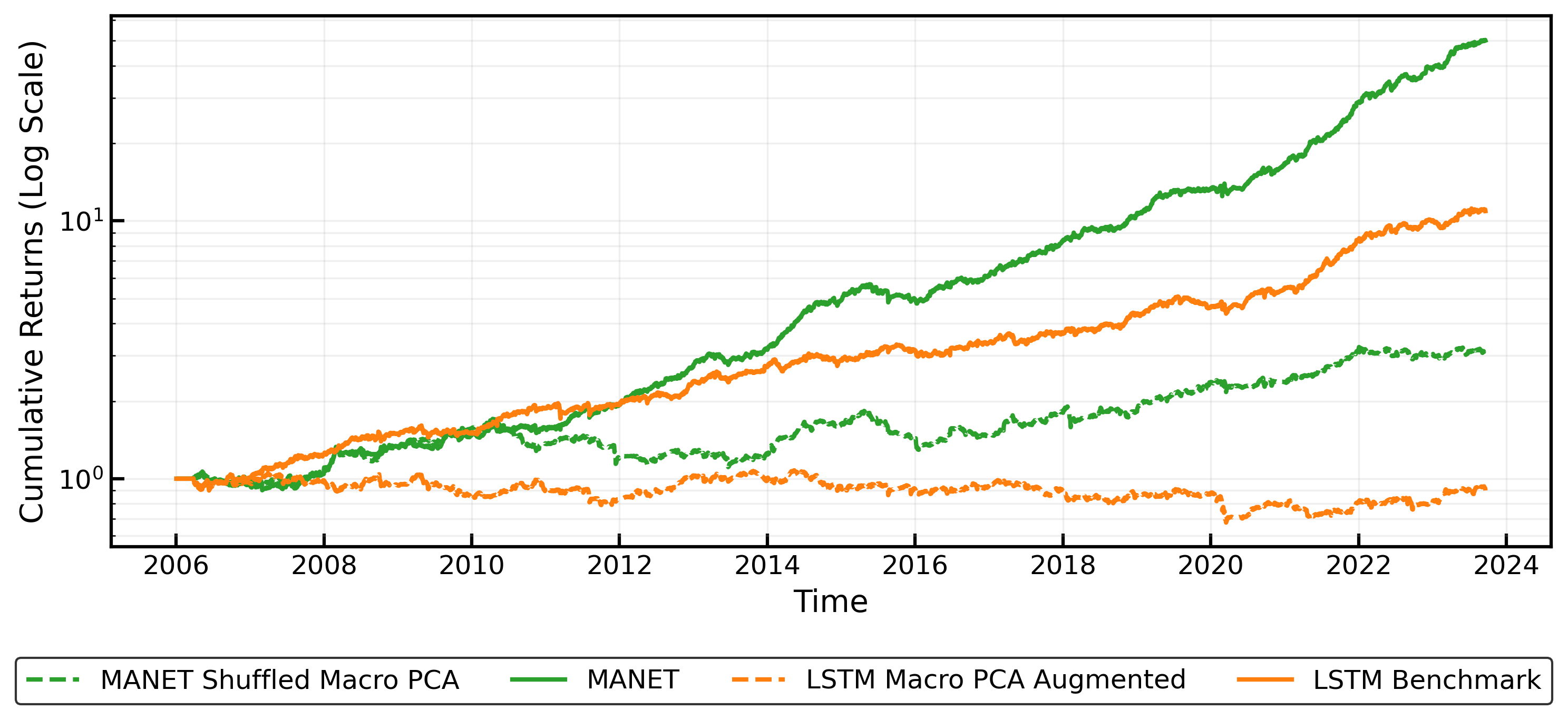}
\caption{Cumulative log-scale returns for the time-series momentum ablation study over the 2005--2024 out-of-sample period. All portfolios are volatility-targeted to 10\% annualized. Solid lines represent the full models (HANET in green, LSTM benchmark in orange), while dashed lines denote their respective ablation variants (HANET with shuffled macro PCA in dashed green, LSTM with concatenated macro PCA in dashed orange). Separation between the solid green and dashed green curves quantifies the contribution of temporal macro structure; separation between the solid green and solid orange curves quantifies the overall benefit of the cross-attention mechanism.}
  \label{fig:tsmom-all-cum-ret-ablation}
\end{figure}

Table~\ref{tab:tsmom-all-portfolio-metrics-ablation} and Figure~\ref{fig:tsmom-all-cum-ret-ablation} further indicate that these gains are specifically tied to \emph{how} macroeconomic information is incorporated, not merely to its inclusion. A naive macro-augmentation baseline, the LSTM with PCA macro inputs concatenated directly to the daily feature vector, performs dramatically worse than the standard LSTM, collapsing to a mean return of just 0.64\% and a Sharpe ratio of 0.06, with a maximum drawdown that nearly doubles ($-25.82\%$ vs.\ $-14.16\%$). This suggests that simply concatenating low-frequency macro factors to a daily forecaster can add noise or create a mismatch in temporal alignment that overwhelms whatever informational benefit the macro data might provide.

In contrast, HANET's hierarchical cross-attention mechanism preserves and amplifies the value of macro information: its Sharpe of 1.99 exceeds even the macro-free LSTM by a wide margin. The shuffled-macro ablation, in which the temporal ordering of the macro PCA factors is randomly permuted (destroying the time-series structure while preserving the marginal distribution), yields a Sharpe of 0.81 and a mean return of 9.41\%. This represents a substantial deterioration relative to the full HANET (a drop in Sharpe from 1.99 to 0.81), confirming that the model is exploiting meaningful temporal patterns in the macro environment rather than benefiting from additional parameters, regularization, or the mere statistical properties of the macro factors. Notably, the shuffled-macro variant still outperforms the naive LSTM+PCA augmentation (0.81 vs.\ 0.06 Sharpe), which underscores that HANET's architecture provides value even when the macro 
signal is degraded, but the full temporal structure is clearly essential for peak performance.

Figure~\ref{fig:tsmom-all-cum-ret-ablation} makes the separation vivid: HANET (solid green) pulls away from all variants by roughly 2014 and sustains a markedly steeper compounding trajectory through the end of the sample. The LSTM benchmark (solid orange) and the shuffled-macro HANET (dashed green) track each other more closely but both lag well behind the full model. The LSTM Macro PCA Augmented variant (dashed orange) flatlines near its starting value throughout the sample, consistent with its near-zero Sharpe ratio, and suffers visibly larger drawdown episodes.

\begin{table}[H]
\centering
\scriptsize
\setlength{\tabcolsep}{10.0pt}
\renewcommand{\arraystretch}{1.05}
\begin{tabular}{l rrrrr}
\toprule
 & \multicolumn{5}{c}{Gross} \\
\cmidrule(lr){2-6}
Strategy & Mean & Std & SR & Sortino & MDD \\
\midrule
LSTM Benchmark & 13.73 & 11.52 & 1.19 & 1.61 & -14.16 \\
LSTM Macro PCA Augmented & 0.64 & 11.46 & 0.06 & 0.07 & -25.82 \\
HANET & 22.45 & 11.31 & 1.99 & 2.50 & -15.87 \\
HANET Shuffled Macro PCA & 9.41 & 11.59 & 0.81 & 0.98 & -16.90 \\
\bottomrule
\end{tabular}
\caption{Ablation results for the time-series momentum task over the 2005--2024 out-of-sample period (gross performance). All portfolios are volatility-targeted to 10\% annualized. The LSTM Macro PCA Augmented variant concatenates the first five principal components of monthly macroeconomic data directly to the daily feature vector, whereas HANET incorporates the same macro factors through hierarchical cross-attention. HANET Shuffled Macro PCA randomly permutes the temporal ordering of the macro principal components, destroying time-series structure while preserving the marginal distribution. Higher mean returns, SR, and Sortino indicate better risk-adjusted performance; a less negative MDD indicates better drawdown control. Comparing across rows isolates the contribution of the cross-attention mechanism and the temporal structure of macro information.}
\label{tab:tsmom-all-portfolio-metrics-ablation}
\end{table}

\begin{table}[H]
\centering
\scriptsize
\setlength{\tabcolsep}{10.0pt}
\renewcommand{\arraystretch}{1.05}
\begin{tabular}{r r r r r}
\toprule
 & \multicolumn{4}{c}{Performance vs.\ Transaction Costs} \\
\cmidrule(lr){2-5}
Cost (bps) & Gross SR & Net SR & $\Delta$ SR & Trades \\
\midrule
0.00  & 1.958 & 1.958 & 0.000  & 4622 \\
0.10  & 1.958 & 1.923 & -0.036 & 4622 \\
0.50  & 1.958 & 1.780 & -0.179 & 4622 \\
1.00  & 1.958 & 1.601 & -0.357 & 4622 \\
2.00  & 1.958 & 1.244 & -0.715 & 4622 \\
5.00  & 1.958 & 0.171 & -1.788 & 4622 \\
10.00 & 1.958 & -1.608 & -3.567 & 4622 \\
\bottomrule
\end{tabular}
\caption{Transaction-cost sensitivity analysis for the volatility-targeted (10\% annualized) HANET time-series momentum portfolio. Each row applies a symmetric per-trade cost (in basis points) to all 4{,}622 round-trip trades executed over the 2005--2024 out-of-sample period. Gross SR is the Sharpe ratio before costs, Net SR is the Sharpe ratio after deducting the assumed cost on each trade, and $\Delta$\,SR reports the resulting deterioration. The table allows readers to assess the strategy's robustness to realistic transaction-cost assumptions: for example, at 1\,bps per trade the net Sharpe ratio remains above 1.60, while costs above 5\,bps eliminate most of the risk-adjusted return.}
\label{tab:tsmom-cost-sensitivity}
\end{table}

\subsection{Time Series Carry Task}

\begin{figure}[H]
  \centering
  \includegraphics[width=0.75\textwidth]{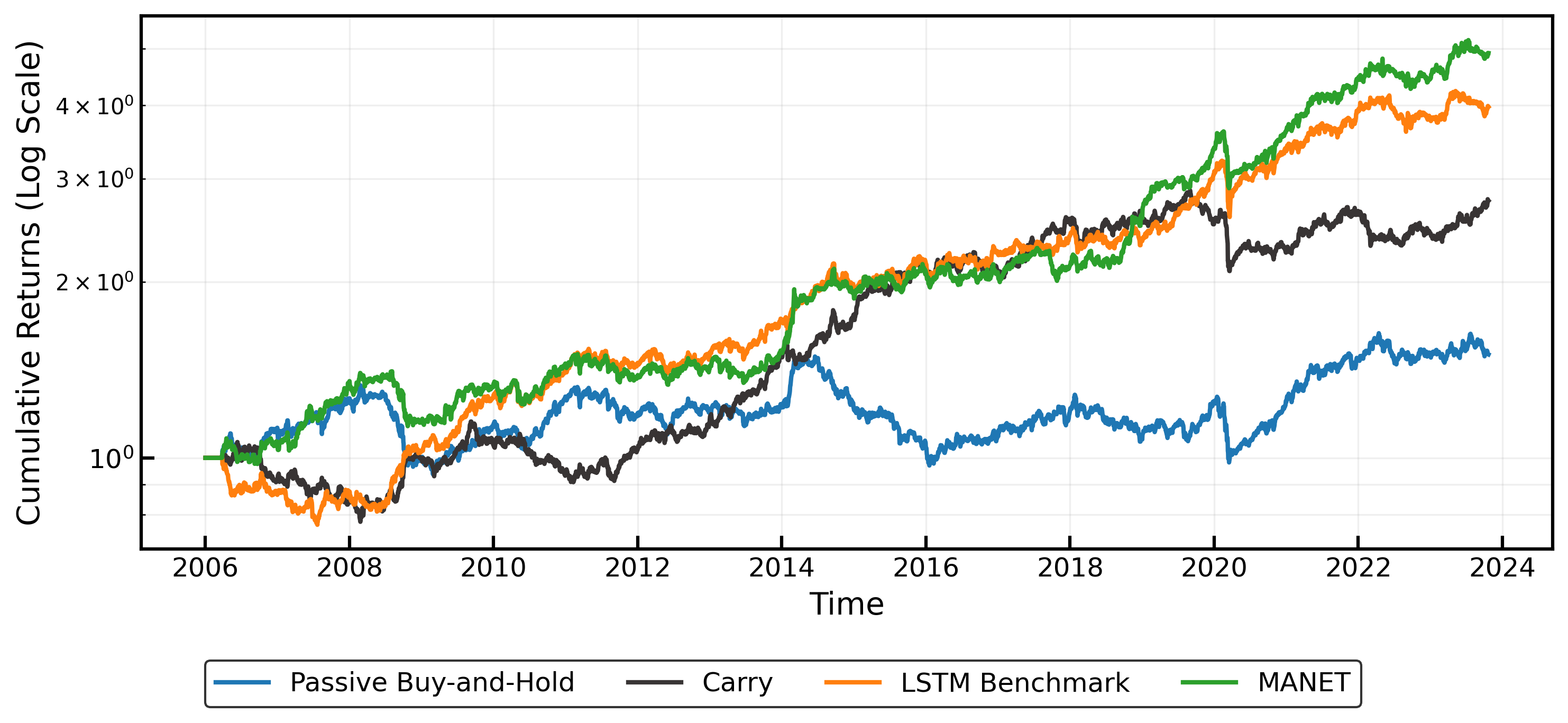}
    \caption{Cumulative log-scale returns (top panel) and drawdowns (bottom panel) for volatility-targeted (10\% annualized) time-series carry portfolios over the 2005--2024 out-of-sample period. Strategies shown are HANET (green), the LSTM benchmark (orange), the classical carry signal (black), and the passive buy-and-hold benchmark (blue). A steeper upward slope in the top panel indicates faster compounding; shallower troughs in the bottom panel indicate milder and shorter-lived drawdown episodes.}
  \label{fig:tscarry-all-cum-ret}
\end{figure}

Table~\ref{tab:carry-portfolio-metrics} and Figure~\ref{fig:tscarry-all-cum-ret} show that HANET also improves carry-portfolio performance, though the magnitudes are more moderate than in the momentum setting. At the same 10\% volatility target (see Section~\ref{sec:ts_factors}), HANET achieves an annualized mean return of 9.49\% with a Sharpe ratio of 0.91 and a Sortino ratio of 1.28, exceeding both the classical carry signal (6.49\% mean, 0.62 Sharpe, 0.93 Sortino) and the LSTM benchmark (7.80\% mean, 0.76 Sharpe, 1.08 Sortino). As in the momentum task, annualized volatilities are tightly clustered across all strategies (10.27\% to 10.49\%), confirming that the improvements reflect better signal quality rather than higher risk-taking. HANET's mean-return $t$-statistic of 3.68 is the highest in the panel and comfortably exceeds conventional significance thresholds.

Second, HANET mitigates crash risk relative to both the classical carry signal and the LSTM benchmark. The passive buy-and-hold benchmark suffers a maximum drawdown of $-34.86\%$, and the classical carry strategy incurs $-27.16\%$, while the LSTM benchmark reduces this to $-21.78\%$. HANET achieves the best drawdown protection in the panel at $-17.71\%$, representing a meaningful improvement over all alternatives. Unlike the momentum task, where the LSTM held a slight edge on drawdowns, here HANET dominates on both return and risk dimensions simultaneously.

Relative to the passive benchmark, HANET exhibits the highest information ratio (0.89) and the strongest active-return $t$-statistic (3.68, HAC-adjusted), confirming that the excess returns are statistically robust. The correlation of HANET's returns with the benchmark ($\rho = 0.65$) is notably higher than in the momentum task, reflecting the fact that carry strategies tend to maintain more persistent directional exposures. The LSTM benchmark shows a moderate correlation ($\rho = 0.52$) and a lower information ratio (0.73), while the classical carry signal exhibits a negative benchmark correlation ($\rho = -0.25$) and a modest IR of 0.24 ($t$-stat of 1.00), indicating that carry-based timing deviates substantially from passive exposure but captures relatively little incremental return on a risk-adjusted basis. HANET's average holding period of 22 days (versus 27 for the LSTM and 30 for carry) is considerably longer than in the momentum setting, consistent with the slower-moving nature of carry signals.

Figure~\ref{fig:tscarry-all-cum-ret} reinforces these findings visually. On a log scale, HANET (green) and the LSTM benchmark (orange) both compound at substantially steeper rates than the classical carry signal (black) and the passive benchmark (blue) throughout the 2005--2024 sample. HANET pulls ahead of the LSTM gradually, with the separation becoming more pronounced after 2016. The classical carry strategy suffers a severe drawdown during the 2008 financial crisis from which it recovers only slowly, while both neural approaches navigate this period with more limited losses. By the end of the sample, HANET's cumulative return on a log scale is visibly above all alternatives.

\begin{sidewaystable}
\centering
\scriptsize
\setlength{\tabcolsep}{8.0pt}
\renewcommand{\arraystretch}{1.05}
\begin{tabular}{l rrrrrrr rrr}
\toprule
 & \multicolumn{7}{c}{Gross} & \multicolumn{3}{c}{vs Bench.\ (Gross)} \\
\cmidrule(lr){2-8}\cmidrule(lr){9-11}
Strategy & Mean & Std & SR & Sortino & MDD & Hold (Days) & $t$-stat$_1$ & $\rho$ & IR & $t$-stat$_2$ \\
\midrule
Passive Buy-and-Hold (Bench.) & 2.48 & 10.49 & 0.24 & 0.32 & -34.86 & 217 & 1.22 & 1.00 & -- & -- \\
\addlinespace[1pt]
Carry & 6.49 & 10.40 & 0.62 & 0.93 & -27.16 & 30 & 2.38 & -0.25 & 0.24 & 1.00 \\
LSTM Benchmark & 7.80 & 10.27 & 0.76 & 1.08 & -21.78 & 27 & 3.31 & 0.52 & 0.73 & 2.99 \\
HANET & \textbf{9.49} & 10.43 & \textbf{0.91} & \textbf{1.28} & \textbf{-17.71} & 22 & \textbf{3.68} & 0.65 & \textbf{0.89} & \textbf{3.68} \\
\bottomrule
\end{tabular}
\caption{Gross performance summary for volatility-targeted (10\% annualized) time-series carry portfolios over the 2005--2024 out-of-sample period. Mean and Std denote annualized mean return and volatility (\%), SR is the annualized Sharpe ratio, Sortino is the annualized Sortino ratio, and MDD is the maximum peak-to-trough drawdown (\%). Hold (Days) reports the average number of trading days a position is maintained before reversal. The right panel reports statistics relative to the passive buy-and-hold benchmark: $\rho$ is the Pearson correlation of daily strategy returns with benchmark returns, IR is the annualized information ratio (mean active return divided by tracking error), and $t$-stat$_2$ is the Newey--West HAC $t$-statistic of active returns. Higher SR, Sortino, and IR indicate superior risk-adjusted performance; a less negative MDD indicates better drawdown control. Best values across strategies are highlighted in bold.}
\label{tab:carry-portfolio-metrics}
\end{sidewaystable}

\begin{table}[H]
\centering
\scriptsize
\setlength{\tabcolsep}{10.0pt}
\renewcommand{\arraystretch}{1.05}
\begin{tabular}{l rrrrr}
\toprule
 & \multicolumn{5}{c}{Gross} \\
\cmidrule(lr){2-6}
Strategy & Mean & Std & SR & Sortino & MDD \\
\midrule
LSTM Benchmark & 7.80 & 10.27 & 0.76 & 1.08 & -21.78 \\
LSTM Macro PCA Augmented & -1.79 & 10.48 & -0.17 & -0.22 & -52.18 \\
HANET & 9.49 & 10.43 & 0.91 & 1.28 & -17.71 \\
HANET Shuffled Macro PCA & 1.24 & 10.64 & 0.12 & 0.15 & -34.59 \\
\bottomrule
\end{tabular}
\caption{Ablation results for the time-series carry task over the 2005--2024 out-of-sample period (gross performance). All portfolios are volatility-targeted to 10\% annualized. The LSTM Macro PCA Augmented variant concatenates the first five principal components of monthly macroeconomic data directly to the daily feature vector, whereas HANET incorporates the same macro factors through hierarchical cross-attention. HANET Shuffled Macro PCA randomly permutes the temporal ordering of the macro principal components, destroying time-series structure while preserving the marginal distribution. Higher mean returns, SR, and Sortino indicate better risk-adjusted performance; a less negative MDD indicates better drawdown control. Comparing across rows isolates the contribution of the cross-attention mechanism and the temporal structure of macro information to carry-based timing.}
\label{tab:tscarry-all-portfolio-metrics-ablation}
\end{table}

\begin{figure}[H]
  \centering
  \includegraphics[width=0.75\textwidth]{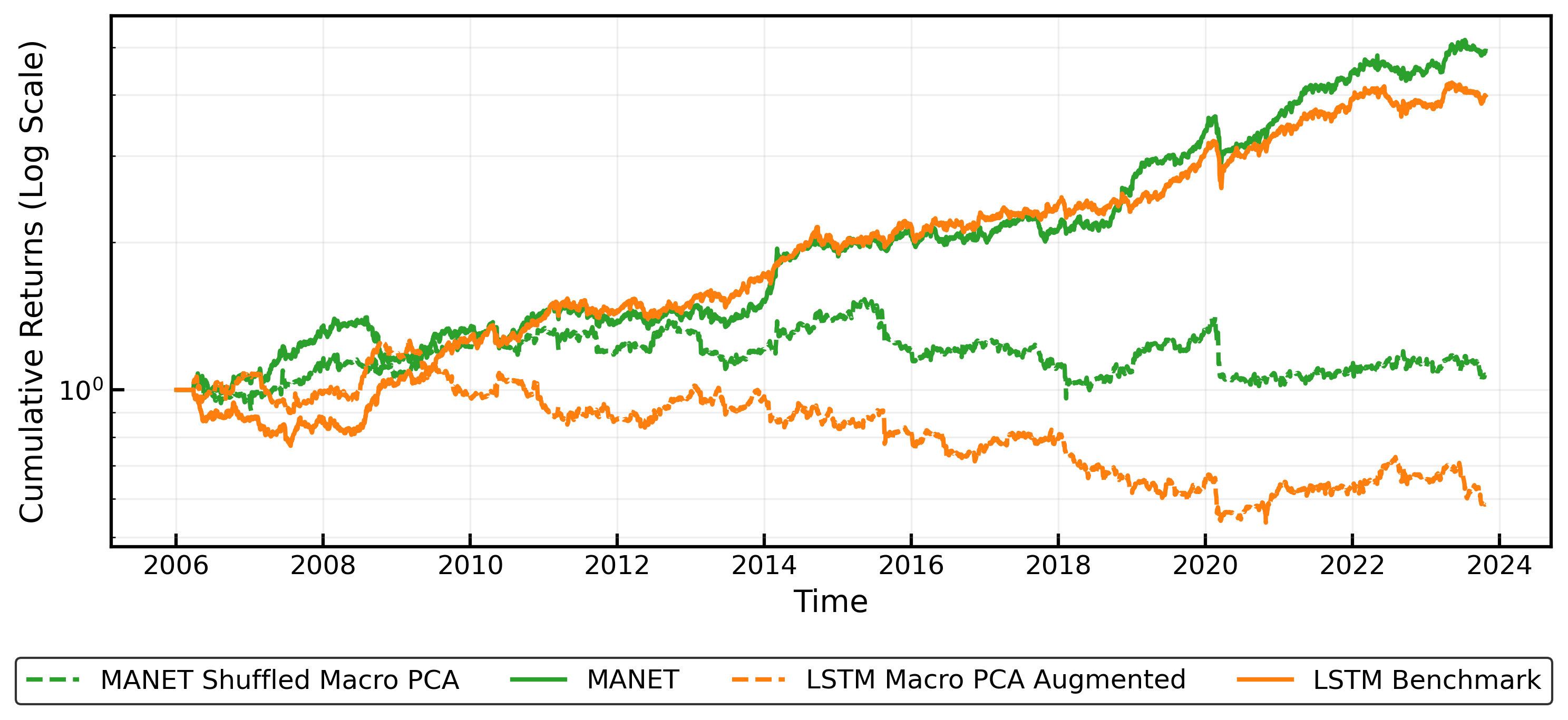}
\caption{Cumulative log-scale returns for the time-series carry ablation study over the 2005--2024 out-of-sample period. All portfolios are volatility-targeted to 10\% annualized. Solid lines represent the full models (HANET in green, LSTM benchmark in orange), while dashed lines denote their respective ablation variants (HANET with shuffled macro PCA in dashed green, LSTM with concatenated macro PCA in dashed orange). Separation between the solid green and dashed green curves quantifies the contribution of temporal macro structure to carry timing; separation between the solid green and solid orange curves quantifies the overall benefit of the cross-attention mechanism.}
  \label{fig:tscarry-all-cum-ret-ablation}
\end{figure}

\begin{table}[H]
\centering
\scriptsize
\setlength{\tabcolsep}{10.0pt}
\begin{tabular}{r r r r r}
\toprule
 & \multicolumn{4}{c}{Performance vs.\ Transaction Costs} \\
\cmidrule(lr){2-5}
Cost (bps) & Gross SR & Net SR & $\Delta$ SR & Trades \\
\midrule
0.00  & 0.913 & 0.913 & 0.000  & 4645 \\
0.10  & 0.913 & 0.910 & -0.003 & 4645 \\
0.50  & 0.913 & 0.897 & -0.016 & 4645 \\
1.00  & 0.913 & 0.881 & -0.031 & 4645 \\
2.00  & 0.913 & 0.850 & -0.063 & 4645 \\
5.00  & 0.913 & 0.756 & -0.156 & 4645 \\
10.00 & 0.913 & 0.599 & -0.313 & 4645 \\
\bottomrule
\end{tabular}
\caption{Transaction-cost sensitivity analysis for the volatility-targeted (10\% annualized) HANET time-series carry portfolio. Each row applies a symmetric per-trade cost (in basis points) to all 4{,}645 round-trip trades executed over the 2005--2024 out-of-sample period. Gross SR is the Sharpe ratio before costs, Net SR is the Sharpe ratio after deducting the assumed cost on each trade, and $\Delta$\,SR reports the resulting deterioration. The carry strategy's longer average holding period (22 days) relative to the momentum task makes it substantially less sensitive to transaction costs: at 2\,bps per trade the net Sharpe ratio remains above 0.85, and even at 10\,bps the strategy retains a net Sharpe of approximately 0.60.}
\label{tab:tscarry-cost-sensitivity}
\end{table}

\subsection{Interpreting Hierarchical Attention Weights in the Macro Space}
\label{sec:interp-macro-att}

\begin{figure}[p]
  \centering
  \includegraphics[width=0.90\textwidth]{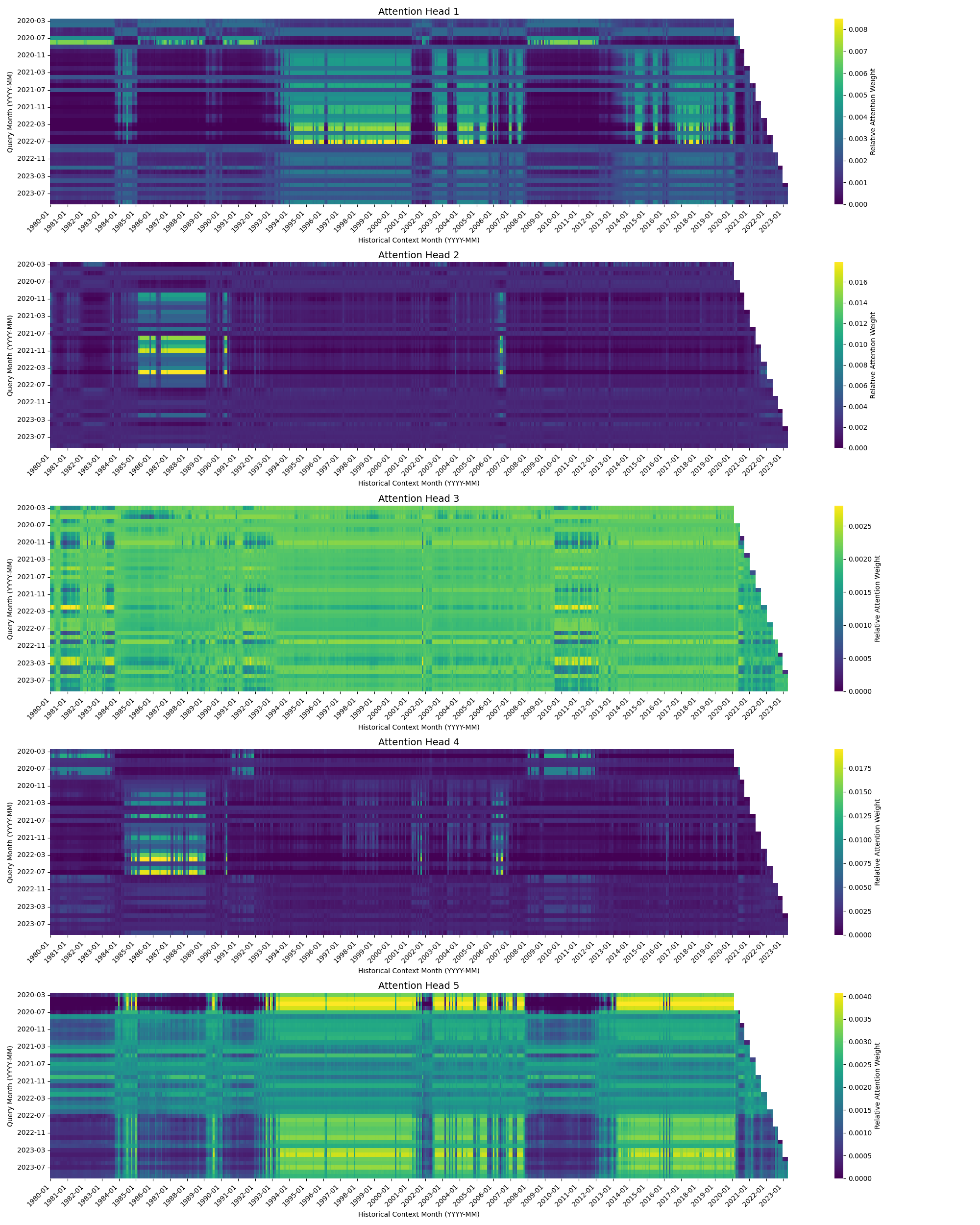}
  \caption{Aggregation of Hierarchical Cross-Attention Weights for the HANET Model in the TSMOM Task. The five stacked heatmaps visualize the attention allocated across the 44-year historical macro-feature matrix (x-axis) for various query months (y-axis). The profiles highlight a functional specialization between sparse ``Regime Detector'' heads (e.g., Heads 2 and 4) and dense ``Global Context'' heads (e.g., Head 5). Relative attention weights are row-normalized to illustrate the model's selective memory across distinct macroeconomic epochs.}
  \label{fig:tsmom-agg-macro-att-weights}
\end{figure}

In this analysis, we showcase the macroeconomic attention weights by utilizing the time-series momentum task, which serves as a rigorous testing ground for the model's ability to relate current price trends to historical macroeconomic environments. To investigate the decision-making process of the HANET architecture, the hierarchical cross-attention weights are visualized to reveal the model's internal ``memory'' logic. Figure~\ref{fig:tsmom-agg-macro-att-weights} presents the aggregate attention profiles across the entire ticker universe, while Appendix~\ref{app:macro-att-selected-futures} details specific profiles for key assets such as the S\&P 500, 10Y US Treasury, Crude Oil, and EURUSD.

Each figure displays one heatmap per attention head. Within each heatmap, the horizontal axis (x-axis) indexes the \emph{historical context month} (YYYY-MM), running from January 1980 on the left to the end of the sample on the right, and represents the pool of past macroeconomic states that the model can attend to. The vertical axis (y-axis) indexes the \emph{query month}, i.e., the date at which the model is generating a forecast; moving from top to bottom corresponds to advancing forward in calendar time through the out-of-sample period. Each row of a heatmap is therefore a single query date's attention distribution over the full historical macro panel, and each cell $(q, c)$ shows how much weight the head places on context month $c$ when producing the forecast for query month $q$. Cells are row-normalized and rendered on a viridis colormap: dark purple indicates negligible weight, teal/green indicates moderate weight, and bright yellow indicates the strongest historical anchors. A \emph{sparse} head concentrates bright cells in narrow vertical bands (the same few historical months are repeatedly retrieved across many query dates), while a \emph{dense} head spreads color more uniformly across the row. A visible diagonal from the top-left to the bottom-right of a panel indicates recency bias, whereby the model preferentially attends to contexts temporally close to the query date.

These visualizations were constructed by computing the sum of attention weights across all tickers in the cross-section for each head, followed by a row-normalization step to ensure the relative importance of each historical month is comparable across different query dates. A critical technical detail of these heatmaps is the inclusion of historical context dating back to 1980. While the futures price data utilized for generating trading signals begins in the 1990s, the underlying macro-feature matrix spans a full 44 years. Context months prior to 1990 are shown despite the absence of ticker-specific returns to provide a complete interpretation of the model's macro-anchoring; they serve as a structural reference that allow the model to link modern query dates to regimes that occurred decades prior.

The attention profiles reveal a sophisticated functional specialization among the five heads, characterized by a clear duality between sparse and dense memory. Heads 2 and 4 operate as ``Regime Detectors'', exhibiting a distinct sparsity bias where high weight is allocated to narrow, vertical bands of history. This indicates a pinpoint focus on specific macro-structural events rather than a general temporal averaging. Conversely, Head 5 provides a dense global context, spreading its attention more uniformly across the 44-year sample to capture long-term cyclical trends and slower-moving economic shifts. This structural duality allows the model to maintain a persistent attention anchor in the early 1980s (the Volcker era) even for query dates in the 2020s. Despite being the most distant context in the "Value" matrix, the model identifies the extreme interest rate volatility of the 1980s as a vital benchmark for navigating current volatile environments.

The interpretability of these weights becomes most evident during global shocks such as the COVID-19 period and the subsequent 2022 inflation cycle. During the COVID-19 shock of 2020, the sparse heads immediately upweighted the historical context of the 2008 Global Financial Crisis. By aligning the 2020 query dates with the 2008 context, the model effectively utilized its long-term memory to navigate a liquidity crisis, recognizing structural similarities in cross-asset correlation spikes despite the differing fundamental causes of the two events. As the query dates moved into the 2022 inflationary spike, HANET demonstrated a remarkable ability to bypass its short-term memory of the ``low-inflation'' 2010s. Instead, it activated its long-term memory to seek context from the early 1980s, the last period characterized by double-digit inflation and aggressive monetary tightening. This dynamic navigation of 44 years of macro-history shed light on the significant performance edge and high HAC $t$-stat$_2$ reported in the primary results, 
suggesting that the model  systematically associates current market conditions with  historically similar macroeconomic environments.

\section{Conclusions}

In this paper, we introduced HANET (Hierarchical Attention Network), a hybrid LSTM-based architecture that systematically integrates macroeconomic information into attention-based deep learning for financial forecasting under scarce and shifting regimes. Our main methodological contribution is a Hierarchical Cross-Attention mechanism for mixed-frequency data: monthly macro contexts define queries and keys, daily market information is preserved in the values, and monthly attention weights are 
projected to the daily grid to reconcile low-frequency macro signals with high-frequency returns. Empirically, across a universe of 55 liquid futures and two systematic tasks (time series momentum and carry), HANET consistently improves risk-adjusted performance relative to classical signals and neural baselines, delivering higher Sharpe/Sortino ratios and smaller drawdowns, with ablations indicating that these gains rely on meaningful macro structure rather than naive macro feature concatenation. These results suggest that framing regime selection as attention over macroeconomic contexts is a promising approach for robust deep learning in finance.

The attention mechanism also provides interpretability. Visualization of the cross-attention weights 
suggests 
functional specialization among heads, with sparse ``regime detector'' heads 
placing greater weight on 
specific historical episodes and dense ``global context'' heads maintaining broad cyclical awareness. During the COVID-19 shock, the model upweighted 2008 crisis contexts; during the 2022 inflationary episode, it 
placed less weight on 
the quiescent 2010s 
and instead assigned greater weight to 
early 1980s Volcker-era contexts. These patterns 
suggest 
that 
HANET systematically associates current market conditions with historically similar macroeconomic environments.

Several avenues follow naturally. First, HANET can be extended to richer macro representations, including nonlinear embeddings (e.g., autoencoders), regime discretizations (e.g., clustering-based states), and multi country macro panels to support global multi-asset trading. Second, the mixed-frequency attention primitive can be generalized beyond monthly--daily to other hierarchies (e.g., weekly--daily, intraday--daily), enabling applications to higher frequency execution settings and event-driven trading. Third, future work should stress test HANET under alternative evaluation protocols, including walk-forward re-estimation, stricter non-overlapping regime splits, and transaction-cost aware objectives, to better characterize robustness under realistic deployment constraints. Fourth, the attention pathway opens a principled route to model diagnostics: one can study stability of retrieved regimes over time, sensitivity to macro revisions, and the impact of missing data or publication lags. Finally, it is promising to integrate HANET's regime retrieval mechanism with end-to-end portfolio learners, such as DeePM \citep{wood2026deepm}. This combination would bridge the current bifurcation between regime-aware forecasting and graph-based policy regularization, yielding architectures that are simultaneously temporally adaptive, via attention, and cross-sectionally robust, via graphs.

\bibliographystyle{chicago}

\bibliography{references}

\clearpage
\begin{appendices}

\section{Dataset Details}
\label{app:dataset_details}

This appendix lists the Pinnacle instruments used in our empirical universe, reported using the original Pinnacle symbol convention. For each instrument we report the Pinnacle ticker, contract name (as defined by Pinnacle), and exchange. In our experiments, trend features are computed from continuous/rolled daily close series, while carry features rely on the term-structure information (front vs.\ deferred contracts for futures; spot and 3-month forwards for FX as described in the main text).

\begin{table}[H]
\centering
\scriptsize
\begin{tabular}{lll}
\toprule
\textbf{Symbol} & \textbf{Contract (Pinnacle MarketName)} & \textbf{Exchange} \\
\midrule
CC & Cocoa & CSC \\
DA & Milk III, Comp & CME \\
GI & Goldman Sachs C.\ I. & IOM(CME) \\
JO & Orange Juice & CSC \\
KC & Coffee & CSC \\
KW & Wheat, KC & CSC \\
LB & Lumber & CSC \\
SB & Sugar \#11 & CSC \\
\midrule
ZC & Corn, Electronic & CBOT \\
ZL & Soybean Oil, Electronic & CBOT \\
ZO & Oats, Electronic & CBOT \\
ZR & Rough Rice, Electronic & CBOT \\
ZW & Wheat, Electronic & CBOT \\
\midrule
ZG & Gold, Electronic & COMEX \\
ZI & Silver, Electronic & COMEX \\
ZK & Copper, Electronic & COMEX \\
ZP & Platinum, Electronic & NYMEX \\
ZA & Palladium, Electronic & NYMEX \\
ZN & Natural Gas, Electronic & NYMEX \\
ZU & Crude Oil, Electronic & NYMEX \\
\midrule
ZF & Feeder Cattle, Electronic & CME \\
ZT & Live Cattle, Electronic & CME \\
ZZ & Lean Hogs, Electronic & CME \\
\bottomrule
\end{tabular}
\caption{Commodity futures used in our universe (Pinnacle tickers, contract names, and exchanges).}
\label{tab:app_pinnacle_commodities}
\end{table}

\begin{table}[H]
\centering
\scriptsize
\begin{tabular}{lll}
\toprule
\textbf{Symbol} & \textbf{Contract (Pinnacle MarketName)} & \textbf{Exchange} \\
\midrule
CB & Canadian 10YR Bond & ME \\
DT & Euro Bond (Bund) & DTB \\
FB & T-Note, 5yr composite & CBOT \\
GS & Gilt, Long Bond & LIFFE \\
TU & T-Notes, 2yr composite & CBOT \\
TY & T-Note, 10yr composite & CBOT \\
UB & Euro Bobl & DTB \\
US & T-Bonds, composite & CBOT \\
UZ & Euro Schatz & DTB \\
\bottomrule
\end{tabular}
\caption{Bond and rates futures used in our universe (Pinnacle tickers, contract names, and exchanges).}
\label{tab:app_pinnacle_bonds}
\end{table}

\begin{table}[H]
\centering
\scriptsize
\begin{tabular}{lll}
\toprule
\textbf{Symbol} & \textbf{Contract (Pinnacle MarketName)} & \textbf{Exchange} \\
\midrule
DX & US Dollar Index & ICE \\
AN & Australian \$\$, composite & IMM(CME) \\
BN & British Pound, composite & IMM(CME) \\
CN & Canadian \$\$, composite & IMM(CME) \\
JN & Japanese Yen, composite & IMM(CME) \\
MP & Mexican Peso, composite & IMM(CME) \\
SN & Swiss Franc, composite & IMM(CME) \\
\bottomrule
\end{tabular}
\caption{FX instruments used in our universe (Pinnacle tickers, contract names, and exchanges).}
\label{tab:app_pinnacle_fx}
\end{table}

\begin{table}[H]
\centering
\scriptsize
\begin{tabular}{lll}
\toprule
\textbf{Symbol} & \textbf{Contract (Pinnacle MarketName)} & \textbf{Exchange} \\
\midrule
ES & Mini S\&P 500 & CME \\
EN & NASDAQ mini & CME \\
ER & Russell 2 mini & CME \\
MD & Mini S\&P 400 & CME \\
YM & Mini Dow Jones (\$5.00) & CBOT \\
\midrule
CA & CAC 40 & MATIF \\
LX & FTSE 100 & LIFFE \\
NK & Nikkei Index & CME \\
XU & Dow Jones EUROSTOXX50 Index & DTB \\
XX & Dow Jones STOXX50 Index & DTB \\
\midrule
FN & Euro, composite & IMM(CME) \\
ZA & Palladium, Electronic & NYMEX \\
\bottomrule
\end{tabular}
\caption{Equity/index futures (and additional selected index-like contracts) used in our universe (Pinnacle tickers, contract names, and exchanges).}
\label{tab:app_pinnacle_equities}
\end{table}

\section{Experimental Configuration}

We considered two model classes in the experiments: a benchmark \textbf{LSTM} and the proposed \textbf{HANET}. For each run, the logged configuration included the dataset name, model name, dropout rate, learning rate, long-only versus long-short output mode, and the main architectural dimensions. For HANET, the printed configuration additionally tracked the query/value LSTM widths, FFNN widths, number of layers, ticker embedding size, and the number of heads in both the hierarchical cross-attention and self-attention blocks. Runtime logs also recorded the active CUDA device and the local GPU environment.

Table~\ref{tab:hanet_vs_lstm_config} summarizes the default experimental configuration used by the script together with the model variants considered.

\begin{table}[H]
\centering
\footnotesize
\setlength{\tabcolsep}{5pt}
\renewcommand{\arraystretch}{0.95}
\begin{tabular}{l l l}
\toprule
\textbf{Parameter} & \textbf{HANET (Proposed)} & \textbf{LSTM Benchmark} \\
\midrule
\textit{Architecture} & & \\
LSTM Hidden Dimension & \{30, 60\} & \{30, 60\} or configured per run \\
Output LSTM Hidden Dimension & \{30, 60\} & -- \\
FFNN Hidden Dimension & \{25, 50, 100\} & -- \\
Output FFNN Hidden Dimension & \{25, 50, 100\} & -- \\
Hierarchical X-Attention Heads & \{2, 5\} & -- \\
Self-Attention Heads & \{2, 5\} & -- \\
Number of Layers & 1 & 1 \\
Ticker Embedding Dimension & \{25, 50, 100\} & \{25, 50, 100\} when enabled \\
Dropout & 0.3 & 0.3 \\
Dropout Type & variational or regular & variational or regular \\
Layer Normalization & True & True \\
Residual Connections & True & -- \\
Initial Attention Temperature & 0.1 & -- \\
Output Activation & \texttt{tanh} (LS) / \texttt{sigmoid} (LO) & \texttt{tanh} (LS) / \texttt{sigmoid} (LO) \\
\midrule
\textit{Optimization} & & \\
Optimizer & Adam & Adam \\
Learning Rate & $10^{-3}$ & $10^{-3}$ \\
Loss Function & Seq-to-seq Sharpe loss & Seq-to-seq Sharpe loss \\
Batch Size & 32 & 32 \\
Gradient Accumulation Steps & 4 & 4 \\
Effective Batch Size & 128 & 128 \\
Max Gradient Norm & 1.0 & 1.0 \\
Mixed Precision & True (CUDA only) & True (CUDA only) \\
\midrule
\textit{Training Config} & & \\
Sequence Length & 84 days & 84 days \\
Days per Context & 21 days & -- \\
Context Months & all available & -- \\
Epochs & 200 & 200 \\
Early Stopping Patience & 20 & 20 \\
Validation Split & 20\% & 20\% \\
Initial In-Sample Window & $252 \times 15$ days & $252 \times 15$ days \\
Walk-Forward Step & $252 \times 5$ days & $252 \times 5$ days \\
Seed Trials & 100 & 100 \\
Top $N$ Ensemble & top 10\% of seeds & top 10\% of seeds \\
\bottomrule
\end{tabular}
\caption{Experimental configuration and parameter choices for the proposed HANET model and the LSTM benchmark.}
\label{tab:hanet_vs_lstm_config}
\end{table}

\section{Hierarchical Attention Weights in the Macro Space for Selected Futures}
\label{app:macro-att-selected-futures}

This appendix presents head-level Hierarchical Cross-Attention weight profiles for four representative futures contracts spanning equities, rates, commodities, and foreign exchange. Each figure displays five heatmaps, one per attention head, where rows index query months in the 2020--2023 evaluation window and columns index the 44-year historical macro-feature context; brighter colors indicate higher row-normalized attention weight. Collectively, these profiles illustrate how HANET allocates macroeconomic context differently across asset classes, with distinct functional specializations emerging across heads.

\begin{figure}[H]
  \centering
  \includegraphics[width=0.95\textwidth]{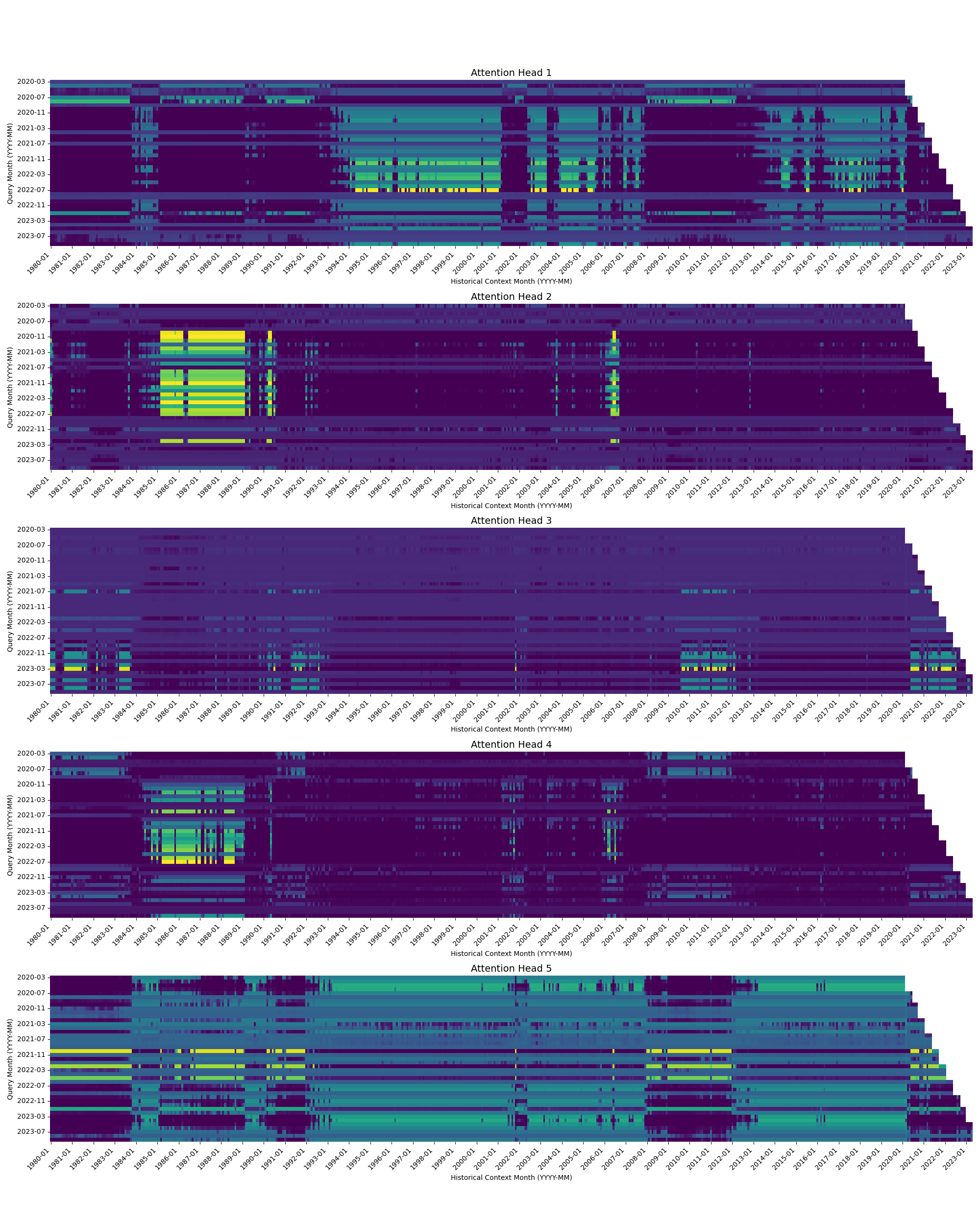}
  \caption{Hierarchical Cross-Attention weight profiles for S\&P 500 futures, by attention head.}
  \label{fig:macro-att-sp500}
\end{figure}

Figure~\ref{fig:macro-att-sp500} shows the attention profile for S\&P 500 futures. Head~1 spreads attention broadly over the post-1995 period but sharpens noticeably during the 2022 inflation cycle, suggesting a partial shift toward historical high-inflation contexts as query conditions evolve. Head~2 behaves as a sparse regime detector, concentrating weight on the mid-1980s and 2008 crisis bands, with visible intensification for 2021--2022 query dates. Head~3 remains largely quiescent across the window, contributing little differential information. Head~4 targets the Volcker era and post-2008 contexts, gradually shifting its anchoring toward the 1980s as 2022 inflationary conditions materialize. Head~5 provides dense global context across the post-2008 period, with a clear diagonal recency structure that traces the model's gradual rolling memory.

\begin{figure}[H]
  \centering
  \includegraphics[width=0.95\textwidth]{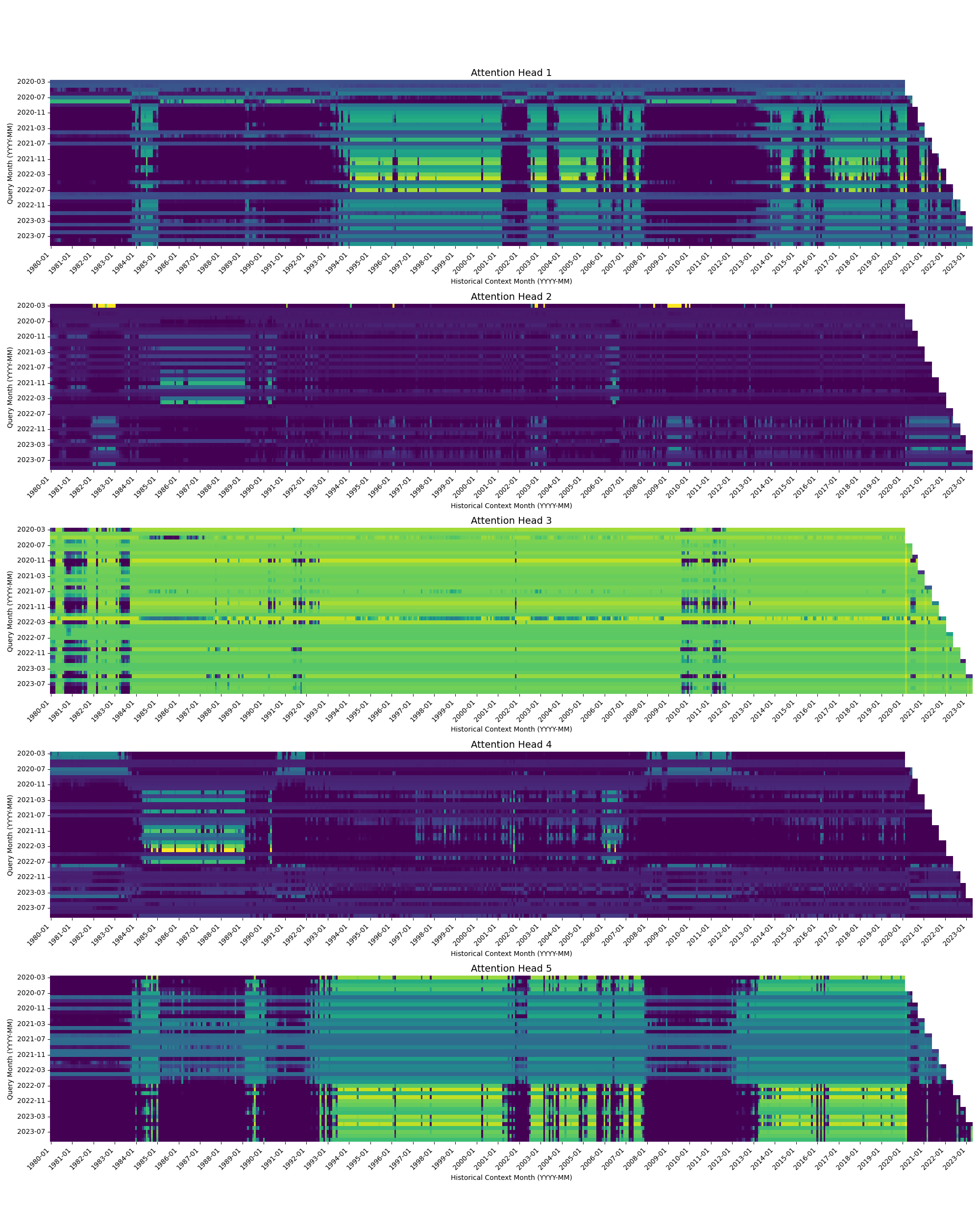}
  \caption{Hierarchical Cross-Attention weight profiles for 10Y US Treasury futures, by attention head.}
  \label{fig:macro-att-ust10y}
\end{figure}

A qualitatively different pattern emerges for 10Y US Treasury futures, shown in Figure~\ref{fig:macro-att-ust10y}. Head~1 distributes broad attention across the entire post-1980 sample, consistent with the long memory inherent in interest rate dynamics. Head~2 acts as a recency gate, concentrating weight on the most recent context months and allocating near-zero attention to pre-2000 periods. Head~3 exhibits sparse, scattered activations that plausibly correspond to the detection of isolated rate-regime transitions. Head~4 provides dense, nearly uniform coverage of the post-2008 low-rate environment, functioning as a persistent context layer for the zero lower bound era. Head~5 shows diffuse attention punctuated by sharp activations around 2008--2009, together with a visible intensification for 2022 query dates that reaches back toward late-1990s tightening episodes.

\begin{figure}[H]
  \centering
  \includegraphics[width=0.95\textwidth]{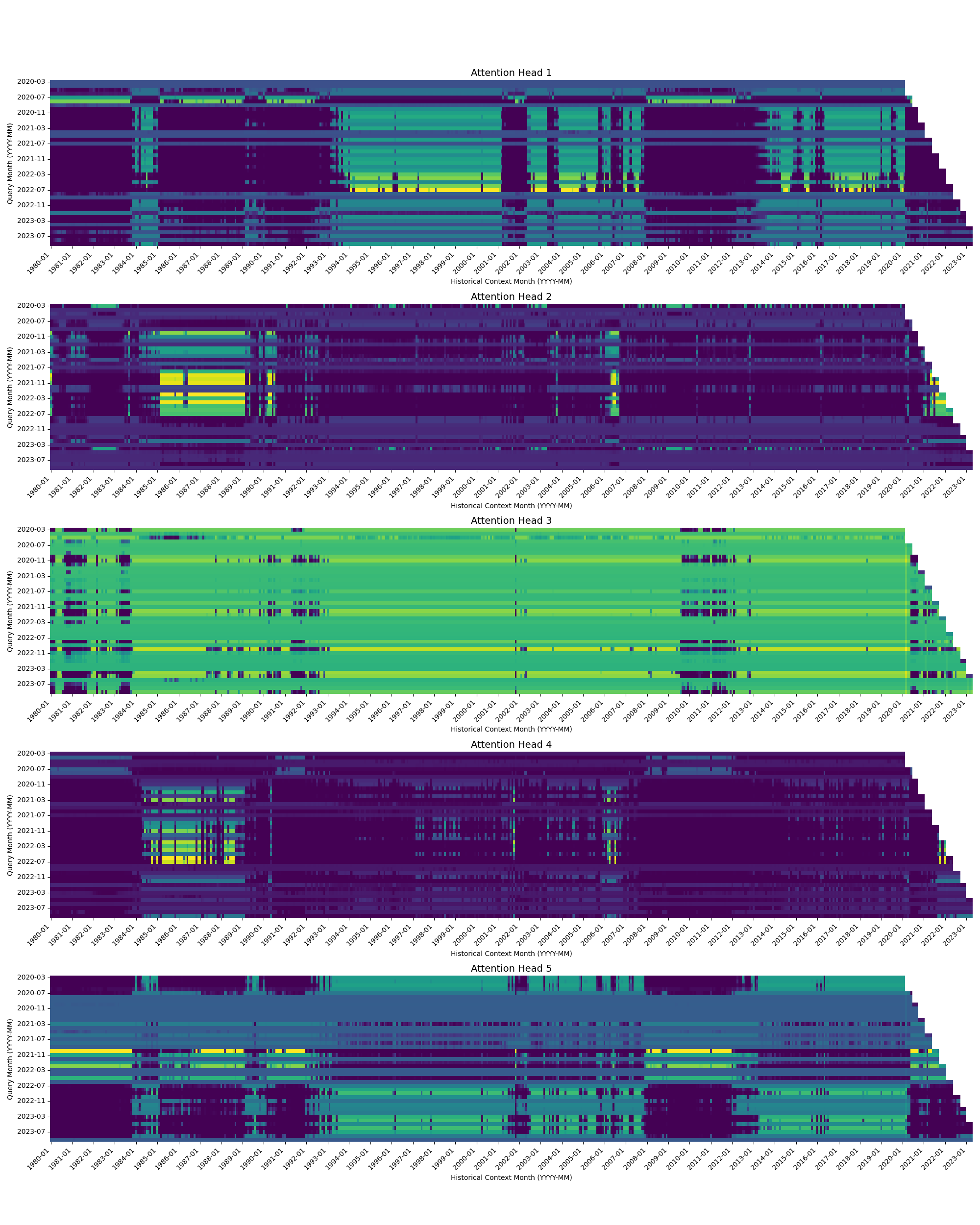}
  \caption{Hierarchical Cross-Attention weight profiles for Crude Oil futures, by attention head.}
  \label{fig:macro-att-crudeoil}
\end{figure}

Turning to commodities, Figure~\ref{fig:macro-att-crudeoil} presents the profile for Crude Oil futures. Head~1 provides broad post-1995 coverage, with marked intensification during COVID-shock query dates. Head~2 again operates as a sparse regime detector, concentrating on the mid-1980s and 2008, and displays sharp activation during 2021--2022 targeting pre-1990 oil supply shock contexts. Head~3 supplies dense, nearly uniform attention from the mid-1980s through the 2010s, functioning as a global cyclical context layer. Head~4 exhibits dense coverage of the post-1990 commodity financialization era. Head~5 combines diffuse post-2000 attention with punctuated 2008 crisis activations and a diagonal recency structure that intensifies during the 2022 energy price spike.

\begin{figure}[H]
  \centering
  \includegraphics[width=0.95\textwidth]{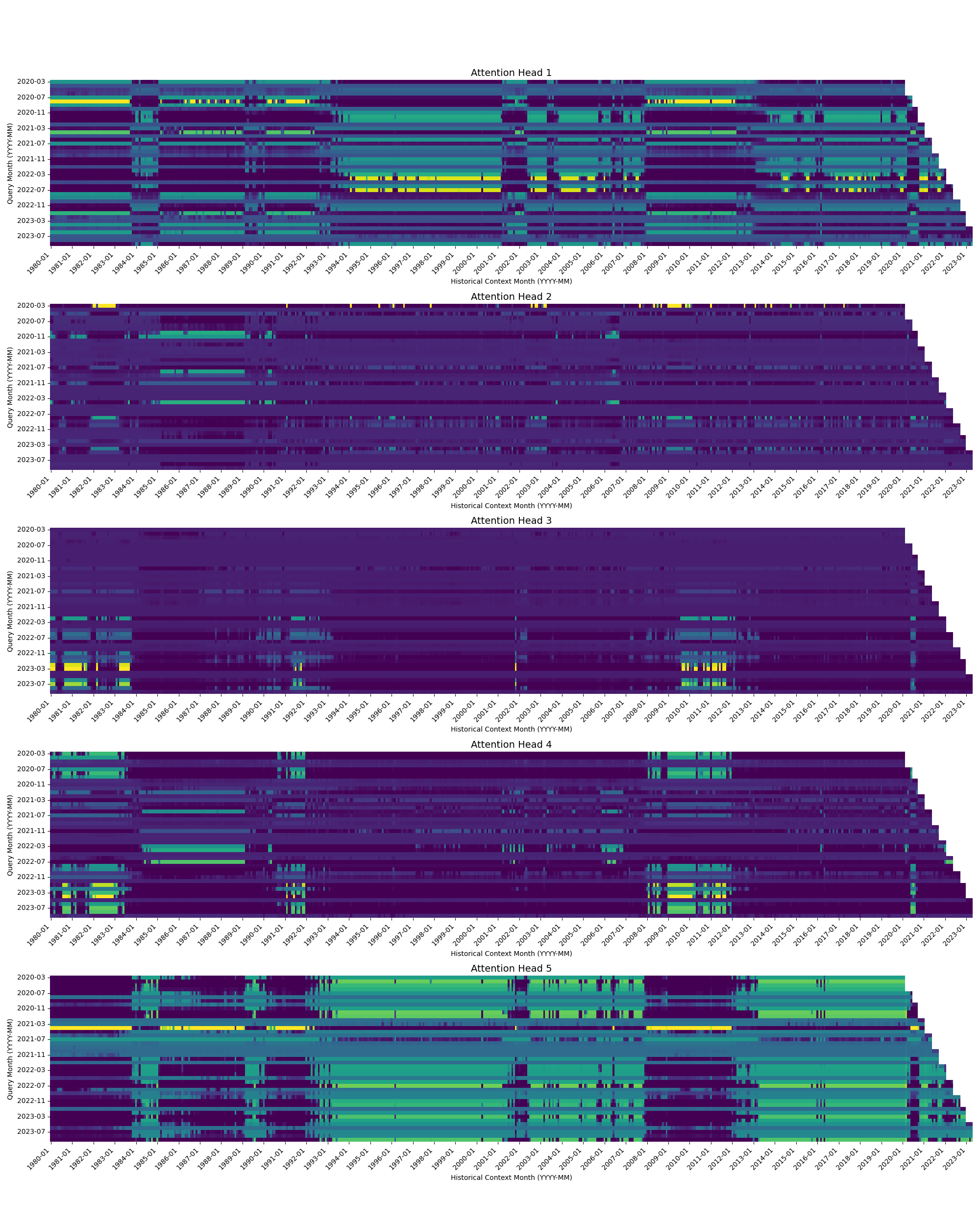}
  \caption{Hierarchical Cross-Attention weight profiles for EURUSD futures, by attention head.}
  \label{fig:macro-att-eurusd}
\end{figure}

Finally, Figure~\ref{fig:macro-att-eurusd} shows the attention profile for EURUSD futures. In contrast to the equity, rates, and commodity profiles, attention patterns here are markedly sparser overall, consistent with the persistent macro regimes---interest rate differentials and relative monetary policy stances---that drive major currency pairs. Head~1 distributes moderate attention over the post-1995 period, with some 2022 intensification. Head~2 is largely quiescent. Heads~3 and~4 exhibit highly sparse activations concentrated on the early-1980s Volcker era, linking the 2022 dollar strengthening episode to the last period of aggressive US tightening. Head~5 provides moderate diffuse coverage of the post-2008 period with mild recency structure.

\end{appendices}

\end{document}